\documentclass[%
 aip,
 amsmath,amssymb,
 reprint,%
]{revtex4-2}

\usepackage{graphicx}%
\usepackage{dcolumn}%
\usepackage{bm}%
\usepackage[hidelinks]{hyperref}%
\usepackage[mathlines]{lineno}%

\usepackage{xcolor}
\usepackage{booktabs}

\usepackage[utf8]{inputenc}
\usepackage[T1]{fontenc}
\usepackage{mathptmx}
\usepackage{etoolbox}

\usepackage{enumitem}

\makeatletter
\def\@email#1#2{%
 \endgroup
 \patchcmd{\titleblock@produce}
  {\frontmatter@RRAPformat}
  {\frontmatter@RRAPformat{\produce@RRAP{*#1\href{mailto:#2}{#2}}}\frontmatter@RRAPformat}
  {}{}
}%
\makeatother
\begin{document}

\preprint{AIP/123-QED}

\title[Diffusion Modeling of Cardiac Waves]{Dreaming of Electrical Waves: Generative Modeling of Cardiac Excitation Waves using Diffusion Models}

\author{Tanish Baranwal}
\affiliation{Cardiovascular Research Institute, University of California, San Francisco, San Francisco, USA}
\affiliation{Department of Electrical Engineering and Computer Science, University of California, Berkeley, USA}

\author{Jan Lebert}
\affiliation{Cardiovascular Research Institute, University of California, San Francisco, San Francisco, USA}

\author{Jan Christoph}
 \email{jan.christoph@ucsf.edu}
 \homepage{http://cardiacvision.ucsf.edu}
 \affiliation{Cardiovascular Research Institute, University of California, San Francisco, San Francisco, USA}
 \affiliation{Division of Cardiology, University of California, San Francisco, San Francisco, USA}

\date{\today}

\begin{abstract}
Electrical waves in the heart form rotating spiral or scroll waves during life-threatening arrhythmias such as atrial or ventricular fibrillation.
The wave dynamics are typically modeled using coupled partial differential equations, which describe reaction-diffusion dynamics in excitable media.
More recently, data-driven generative modeling has emerged as an alternative to generate spatio-temporal patterns in physical and biological systems.
Here, we explore denoising diffusion probabilistic models for the generative modeling of electrical wave patterns in cardiac tissue.
We trained diffusion models with simulated electrical wave patterns to be able to generate such wave patterns in unconditional and conditional generation tasks.
{For instance, we explored the i) parameter-specific diffusion-based generation, ii) evolution and iii) inpainting of spiral wave dynamics, including reconstructing three-dimensional scroll wave dynamics from superficial two-dimensional measurements.
Further, we generated arbitrarily shaped bi-ventricular geometries and simultaneously initiated scroll wave patterns inside these geometries using diffusion.}
We characterized and compared the diffusion-generated solutions to solutions obtained with corresponding biophysical models and found that diffusion models learn to replicate spiral and scroll waves dynamics so well that they could be used for data-driven modeling of excitation waves in cardiac tissue.
{For instance, an ensemble of diffusion-generated spiral wave dynamics exhibits similar self-termination statistics as the corresponding ensemble simulated with a biophysical model.}
However, we also found that diffusion models {produce artifacts if training data is lacking, e.g. during self-termination,} and `hallucinate' wave patterns when insufficiently constrained.
\end{abstract}

\maketitle

\section{Introduction}
\label{sec:introduction}

Waves in excitable media exhibit complex spatio-temporal dynamics \cite{Krinsky1991,Winfree1994}.
In two-dimensional media, they form linear, focal or rotating spiral-shaped waves or compositions thereof. 
In three-dimensional media, they manifest as planar or spherical focal waves, or, if perturbed, %
take on more complicated rotational shapes referred to as scroll waves.
Spiral and scroll wave dynamics have been studied for many decades, as they are associated with %
heart rhythm disorders,
such as atrial fibrillation, polymorphic ventricular tachycardia, or ventricular fibrillation \cite{Winfree1994,Alonso2016,Rappel2022,Pertsov1993,Gray1995,Pathmanathan2015,Christoph2018,Uzelac2022,Fenton1998,Berenfeld1999,Qu2000b}.
In the heart, electrical excitation initiates the contraction of the heart muscle and it is hypothesized that the abnormal, rapid and irregular contractions during cardiac tachyarrhythmias are caused by spiral- and scroll-shaped waves of electrical excitation.

The electrical waves can be reproduced and studied in computer simulations using biophysical models  \cite{FitzHugh1961,Nagumo1962,AlievPanfilov1996}. 
These models %
consist of coupled partial differential equations (PDEs), which describe the electrical excitability $u$ and refractoriness $r$ of cardiac muscle cells and the coupling between them, see eqs.~(\ref{eq:modelu}--\ref{eq:modelr}).
The equations %
model reaction-diffusion dynamics, where the exchange of currents through ion channels between cells are modeled as a diffusive process and the cells as nonlinear oscillators.
Integrating these equations in time and over space in a spatially extended system using, for instance, the finite difference or finite element method produces nonlinear waves of electrical excitation mediated via diffusion.

\begin{figure}[htb]
  \centering
  \includegraphics[width=0.48\textwidth]{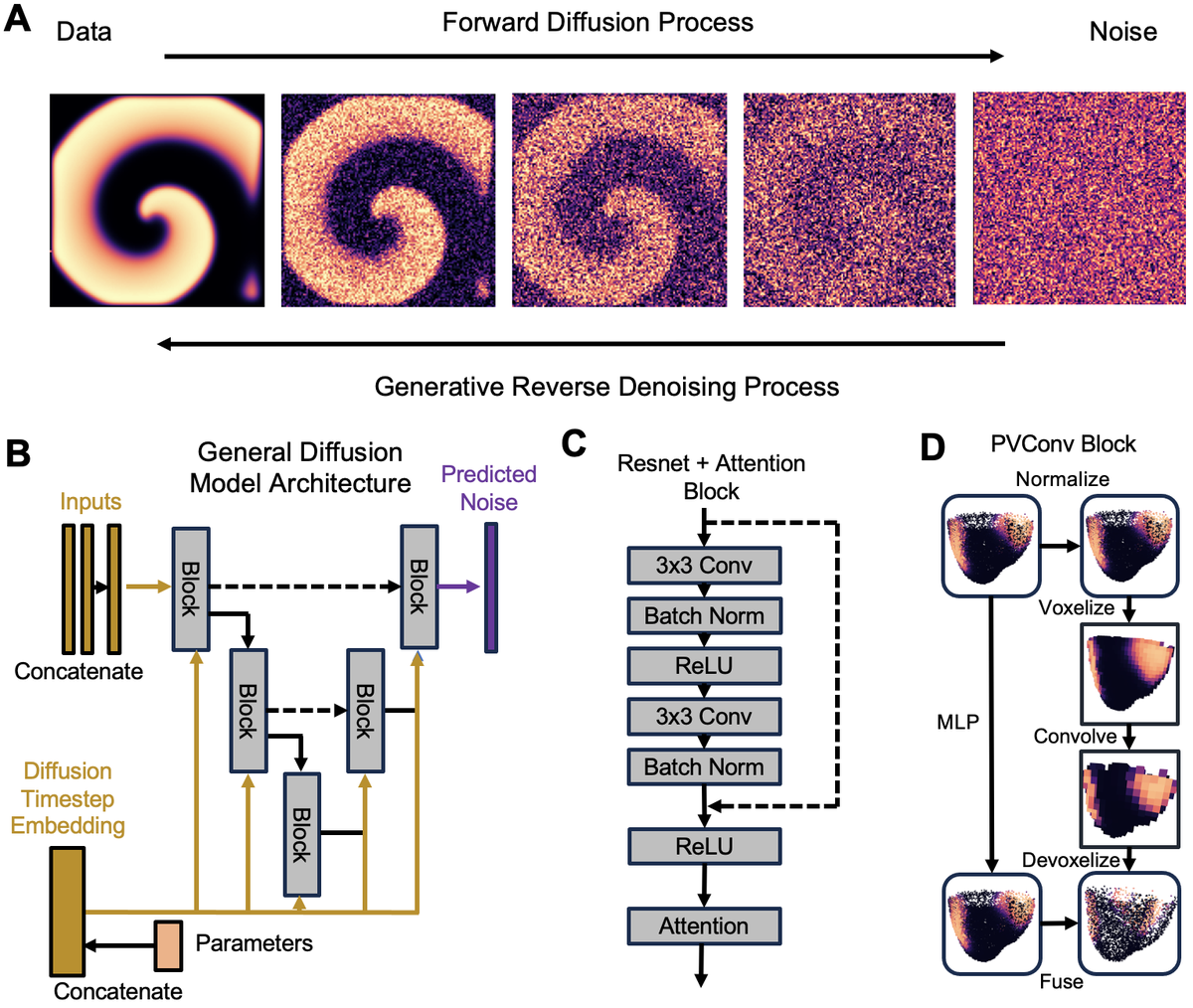}
  \caption{
Diffusion-based generative modeling of electrical wave dynamics in cardiac tissue.
\textbf{A} Forward diffusion process and generative reverse denoising process. The training data consists of spiral and scroll wave dynamics in excitable media.
\textbf{B} General diffusion model architecture for processing image data with underlying U-Net architecture.
\textbf{C} ResNet Attention block.
\textbf{D} Diffusion model for generating scroll waves in heart-shaped geometries represented as pointclouds with corresponding scalar-valued data (Point-Voxel Diffusion \cite{Zhou2021}).
  }
  \label{fig:Fig1}
\end{figure}

Diffusion, on the other hand, is a term that has recently emerged in the field of artificial intelligence (AI), referring to a class of generative neural networks which employ a diffusive process to generate data \cite{Sohl-Dickstein2015,Song2019,Ho2020}.
During the training procedure, noise is iteratively added to the training data and the neural networks, termed denoising diffusion probabilistic models (DDPMs) \cite{Ho2020} or diffusion models, learn to reverse this process, ultimately enabling them to create data from noise, see Fig.~\ref{fig:Fig1}. 
Diffusion models are very successful in generating data such as images \cite{Rombach2021,Saharia2021,Saharia2022}, videos \cite{Singer2022}, and audio \cite{Kong2021}, and they are increasingly also used for technical applications in physics, engineering, medicine, and biology \cite{Pinaya2022,Watson2022,Guo2023,Kazerouni2023}.
Diffusion models likely also have many useful applications in cardiology that yet have to be explored.
For example, they could be used in electrophysiological studies to generate synthetic action potential wave patterns and arrhythmia morphologies, either to fill in or reconstruct missing measurement data, or to simulate cardiac dynamics in a data-driven fashion.
Diffusion-generated solutions could be particularly useful in situations in which measurements can only be obtained partially or indirectly, or when biophysical model equations or parameters are lacking.

In this numerical study, we explore diffusion models for their application in cardiac electrophysiology and arrhythmia research.
We investigated whether they can be used to reconstruct or simulate electrical impulse phenomena in computer simulations of excitable media,
and simulated electrical spiral and scroll waves in two- and three-dimensional square-, bulk- and heart-shaped tissues with isotropic and anisotropic diffusive spread of the excitation.
{
More specifically, we used diffusion models for the following tasks:
\begin{enumerate}[label=\textit{Task \arabic*:}, leftmargin=*, itemsep=-1ex]
  \item Generation of parameter-specific two-dimensional spiral waves, see section~\ref{sec:results:parameterspecific}.
  \item Generation of scroll waves in bi-ventricular heart-shapes, see section~\ref{sec:results:3Dheart}.
  \item Prediction of the evolution of spiral wave dynamics over time, see section~\ref{sec:results:evolve}.
  \item Reconstruction of three-dimensional scroll waves from two-dimensional surface observations, see section~\ref{sec:results:3D}.
  \item Inpainting of two-dimensional spiral wave dynamics, see section~\ref{sec:results:2Dinpainting}.
  \item Unconditional generation of two-dimensional spiral wave patterns, see section~\ref{sec:results:similarity}.
\end{enumerate}
}
We determined how reliable diffusion models are when generating such spatio-temporal physiological dynamics.
Generative neural networks, such as diffusion models, generative adversarial networks (GANs), or large language models (LLMs) are known to be capable of producing a continuum of output including false or undesired output, which is often referred to as 'hallucination'. %
We show that diffusion models can generate electrical waves in many different ways: out of the blue in an unconstrained generative process or when the generative process is constrained or guided by parameters or boundary conditions such as partial data, or a recent dynamical state of the system.
In particular, the latter generative mode corresponds to diffusion-based data-driven modeling of cardiac dynamics.
We found that hallucination occurs when the generation task is insufficiently constrained, which raises concerns over the reliability of diffusion models in diagnostic applications.

\begin{figure*}[htb]
  \centering
  \includegraphics[width=0.98\textwidth]{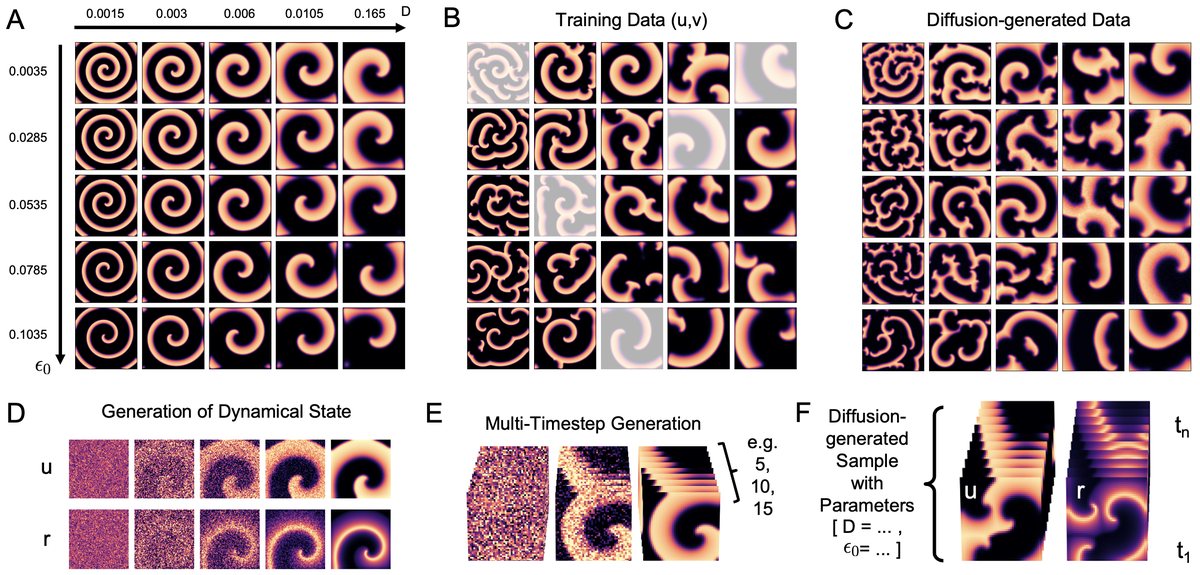}
  \caption{
Parameter-specific generation of spiral wave dynamics using diffusion-based generative modeling (Task 1), see also Supplementary Videos 1-5.
\textbf{A} {Different simulations of spiral waves while varying parameters $D$ and $\epsilon_0$ in eqs.~(\ref{eq:modelu})-(\ref{eq:modelr}), or diffusion constant and time scale separation parameter, respectively, which influence the width of and distance between the waves, respectively.}
\textbf{B} {The diffusion model was trained with data consisting of multi-spiral wave dynamics for the same parameter combinations ($D$,$\epsilon_0$) as in A with $500$ simulations per combination. Some parameter combinations (white) were left out during training (5 of 20).}
\textbf{C} {The diffusion model generates parameter-specific spiral wave patterns for all parameter combinations, even though it was not trained on all of them.
\textbf{D} The diffusion model can generate a full dynamical state with both dynamic variables $u$ and $r$ as well as \textbf{E} multiple timesteps of such states at once: $(u,r)(x,y,t_1, ... , t_n)$ with $t_n = 2,3, ... , 15$.
\textbf{F} Diffusion-generated multi-timestep sample ($t_n = 10$) showing spatio-temporal spiral wave pattern with fast and slow variables (corresponds to $t_s = 100$ simulation time steps, see section \ref{sec:results:parameterspecific}).}
  }
  \label{fig:parameter-specific1}
\end{figure*}

\section{Methods}
\label{sec:methods}

\subsection{Simulations of Electrical Wave Dynamics in Heart Muscle Tissue}
\label{sec:methods:simulations}

We simulated nonlinear waves of electrical excitation using coupled partial differential equations (PDEs) in i) two-dimensional rectangular-shaped, ii) three-dimensional bulk-shaped, and iii) three-dimensional heart-shaped geometries, respectively.
In all three cases, we used the phenomenological Aliev-Panfilov model \cite{AlievPanfilov1996}: 
\begin{eqnarray} 
\label{eq:modelu}
\frac{\partial u}{\partial t} & = & \nabla \cdot (\bm{D} \nabla u) - k u (u-a) (u-1) - u r \\
\label{eq:modelr}
\frac{\partial r}{\partial t} & = & \left(\epsilon_0 + \frac{\mu_1 r}{u+\mu_2}\right) (k u(a+1-u)-r)
\end{eqnarray}
The dynamic variables $u$ and $r$ represent the local electrical excitation and refractoriness in dimensionless, normalized units, respectively.  
The parameters $D$, $k$, $a$, $\epsilon_0$, $\mu_1$ and $\mu_2$ determine the properties of the waves (e.g. excitability, wavelength, conduction speed / {diffusivity}, number of waves {and distance between them}, etc.).
We varied the parameters {$D$} and $\epsilon_0$ to change the properties of the excitation waves and produce different training data for different tasks (Task 1-6), see Table~\ref{tab:parameters} and sections \ref{sec:results:parameterspecific}-\ref{sec:results:transients}.
The simulations in the simplified (rectangular, bulk) and heart-shaped geometries were performed as described in \citet{Lebert2023,Lebert2023b}, respectively.
Correspondingly, the system of equations (\ref{eq:modelu}-\ref{eq:modelr}) was integrated using the forward Euler method and the smoothed particle hydrodynamics method \cite{Zhang2021,Zhang2021b}, respectively.
{All simulations were performed in dimensionless units with $\Delta x = 1$ and integration time steps which proved to be numerically stable.} 
The two-dimensional simulations were isotropic, whereas the three-dimensional simulations were anisotropic with a locally varying fiber direction and faster wave propagation along the fiber direction, {see Table \ref{tab:parameters}.}
The fiber architectures were created as described in \citet{Lebert2023,Lebert2023b}.
The bi-ventricular heart geometries and underlying rule-based fiber architectures were randomly initialized.

The simulation/model parameters were chosen specifically for each task, see Table \ref{tab:parameters}.
For example, we simulated a range of parameter-specific spiral wave dynamics (Task 1), as seen in Fig.~\ref{fig:parameter-specific1}A,B), by varying the parameters {$D$} and $\epsilon_0$ {and by applying a random number of pacing stimuli applied in random locations to cause wave break and create spiral waves}.
We simulated two different regimes of spiral wave dynamics, as shown in Figs.~\ref{fig:evolution} and \ref{fig:hallucination}, using two different parameter sets: one with few (Task 3a, 5a) and one with more spiral waves (Tasks 3b, 5b).
{We simulated scroll wave dynamics in a bulk with $128 \times 128 \times 40$ voxels as shown in Fig.~\ref{fig:3D1} (Task 4) and in bi-ventricular geometries as shown in Fig.~\ref{fig:3Dheart} and described in Lebert et al. \cite{Lebert2023} (Task 2) using a fixed set of parameters.}
For each task, we performed hundreds of simulations to generate sufficient training data and separated training data and data used for evaluation.
For example, for Task 4, we performed $125$ simulations, where $100$ simulations were used for training and $25$ for evaluation, as described in \citet{Lebert2023}.
The initial conditions $u_0, r_0$ were randomized and therefore different in each simulation, see also \citet{Lebert2023b}.
If the spiral or scroll wave dynamics self-terminated prematurely, we restarted the simulation.

Using the simulation data, we generated different training datasets for each task, see Tasks 1-6 in Table \ref{tab:parameters} and sections  \ref{sec:results:parameterspecific}-\ref{sec:results:transients} for details.
{We found that the diffusion models discussed in sections \ref{sec:results:parameterspecific} and \ref{sec:results:similarity} can already generate spiral wave patterns with as few as $100$ training samples, see Supplementary Fig.~2. However, in order to increase the diversity and quality of the generations, we typically used thousands to tens of thousands of training samples for all Tasks 1-6, see Table \ref{tab:training_dataset_size}.}
Each training dataset consisted of samples randomly chosen from only the different training simulations. 
Correspondingly, each evaluation dataset consisted of samples randomly chosen from only the evaluation simulations.
Training and evaluation datasets {were sampled from} completely separate datasets. {There is no overlap or cross-talk between training and evaluation samples.}

\begin{table}[htb]
  \begin{tabular}{@{}c|cccccc@{}}\toprule
    Param.    & Task  & Task & Task & Task & Task & Task \\
                       & 1  & 2 & 3a/5a & 3b/5b & 4 & 6 \\ \midrule
    $D$          & {$D'$}   & {$\textbf{D}$} & 1 & 1 &  {$\textbf{D}$} & 1 \\
    $k$          &  8    & $8$  &$8.5$ & $7.5$ & $8$ & $8$    \\
    $a$          & 0.1  & $0.2$ &$0.1$ & $0.1$ & $0.05$ & 0.05 \\
    $\epsilon_0$  & $\epsilon_0'$  & $0.002$  &$0.003$ & $0.001$ & $0.002$ & $0.002$ \\
    $\mu_1$      & 0.2   & $0.2$   &$0.16$ & $0.16$ & $0.8$ &0.2 \\
    $\mu_2$      & 0.3    & $0.3$  &$0.3$& $0.3$ & $0.3$ &0.3  \\\bottomrule
  \end{tabular}
  \caption{
    Parameters of biophysical model \cite{AlievPanfilov1996} used to simulate electrical wave patterns. 
    Task 1: Parameter-specific generation of spiral waves, see Figs.~\ref{fig:parameter-specific1} and \ref{fig:parameter-specific2}.
    Task 2: Scroll wave dynamics in bi-ventricular heart-shaped medium, see Fig.~\ref{fig:3Dheart}.
    {
    $\textbf{D}$ is an anisotropc diffusion tensor, see \citet{Lebert2023b}.
    The diffusion coefficient $D_{\perp}$ perpendicular to the fiber orientation was set to $0.1$~mm$^2$/ms, the parallel coefficient $D_{\parallel}$ was randomly chosen for each simulation from the interval $0.2-0.4$~mm$^2$/ms.
    }
    Tasks 3, 5, 6: Spiral wave dynamics in 2D isotropic medium, see Figs.~\ref{fig:evolution}, \ref{fig:hallucination}, \ref{fig:2Dclassification}.
    Task 4: Scroll waves in anisotropic 3D bulk shown in Fig.~\ref{fig:3D1}.
    {
      $\textbf{D}$ is an anisotropc diffusion tensor with $D_{\perp}=0.05$ and $D_{\parallel}=0.2$, see \citet{Lebert2023} for details.
    }
    }
  \label{tab:parameters}
\end{table}

\begin{figure*}[htb]
  \centering
  \includegraphics[width=0.98\textwidth]{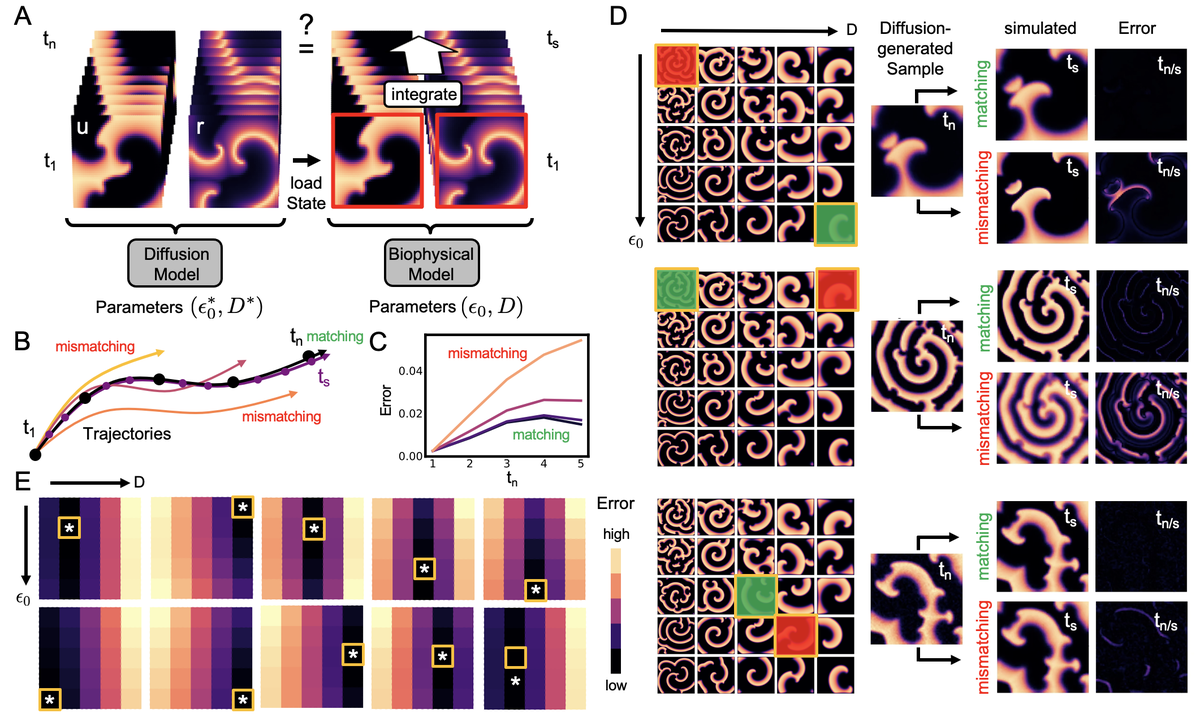}
  \caption{
{Scheme to verify whether the diffusion-generated spiral wave patterns in Fig.~\ref{fig:parameter-specific1} are parameter-specific:
\textbf{A} The first state $(u,r)$ at $t_1$ of the diffusion-generated multi-timestep sample generated with parameters $(\epsilon_0^*,D^*)$ was loaded into the corresponding biophysical model with either identical or mismatching parameters $(\epsilon_0,D)$. 
The biophysical model was then integrated for $t_s$ time steps and the solutions were then compared to the last state of the diffusion-generated sample at $t_n$. 
This was repeated for all parameter combinations $(\epsilon_0,D)$, see section \ref{sec:results:parameterspecific} for details.
\textbf{B} Trajectories in phase space starting from state $t_1$, co-evolving with matching parameters and diverging with mismatching parameters.
\textbf{C} Error between diffusion-generated and simulated states over time ($t_n = 5$ corresponds to $t_s=150$) with matching (low error: black/purple) and mismatching (high error: orange/yellow) parameters for one diffusion-generated sample.
\textbf{D} Examples of diffusion-generated and simulated states at $t_n = t_s$ with matching parameters ($(\epsilon_0^*,D^*)=(\epsilon_0,D)$) and mismatching parameters ($5 \times 5$ grid with same parameter combinations as in Fig.~\ref{fig:parameter-specific1}A,B).
The diffusion-generated sample was generated with the parameter combination indicated by the green square. The matching biophysical simulation was performed with the same parameters, while the mismatching simulation was performed with the combination indicated by the red square.
The simulations deviate from the diffusion-generated samples when the parameters do not match which causes an error (pixel-wise absolute difference).
\textbf{E} Confirmation that generations are parameter-specific: the average pixel-wise error (MAE) between simulated and diffusion-generated patterns at $t_n$ (averaged over 100 samples/simulation) is the lowest (black/purple: small, orange/yellow: large) for matching parameter combinations $(D,\epsilon_0) = (D^*,\epsilon_0^*)$. 
$\square$ parameter combination $(D^*,\epsilon_0^*)$ used to generate diffusion-generated sample; $*$ parameter combination of simulation $(D,\epsilon_0)$ with lowest error; $\square$ and $*$ match in 21 out of 25 cases, in the other cases the minimum is nearby with marginal difference in the error. $5 \times 5$ grid corresponds to same parameter combinations as in Fig.~\ref{fig:parameter-specific1}A.}
  }
  \label{fig:parameter-specific2}
\end{figure*}

\subsection{Denoising Diffusion Model}
\label{sec:methods:diffusionmodel}

We used a denoising diffusion probabilistic modeling \cite{Ho2020} neural network architecture, which we refer to as diffusion model for simplicity.
Diffusion models consist of a forward diffusion process and a reverse diffusion process, see Fig.~\ref{fig:Fig1}.
During the forward diffusion process, gaussian noise is added incrementally to an input image until it is indistinguishable from random noise. 
This produces a sequence of samples $(x_0, \ldots, x_T)$ with increasing noise, starting from the data point $x_0$ from the real data distribution $q(x)$ and ending with what is indistinguishable from an isotropic Gaussian distribution. 
\begin{eqnarray} 
\label{eq:diffusion}
q (\mathbf{x}_t|\mathbf{x}_{t-1})  = \mathbf{N} (\mathbf{x}_t; \sqrt{1-\beta_t}\mathbf{x}_{t-1}, \beta_t \mathbf{I})
\end{eqnarray}
The step sizes are controlled by a variance schedule $\beta_t$. 
\begin{eqnarray}
\label{eq:diffusion1}
q ( \mathbf{x}_{1:T} | \mathbf{x}_0) = \prod^{T}_{t=1}q(\mathbf{x}_t|\mathbf{x}_{t-1})
\end{eqnarray}
When sampling new data from the data distribution $q(x)$, a model $p_{\theta}$ is learned to estimate $q(x_{t-1}|x_t)$, which is also approximated by a gaussian distribution. This is referred to as the reverse diffusion process.
\begin{eqnarray} 
\label{eq:diffusion2}
p_\theta ( x_{0:T}) = p(\mathbf{x}_T) \prod_{t=1}^{T} p_\theta(\mathbf{x}_{t-1}|\mathbf{x}_t)\\
\label{eq:diffusion3}
p_\theta(\mathbf{x}_{t-1}|\mathbf{x}_t) = \mathbf{N}(\mathbf{x}_{t-1};\mathbf{\mu}_\theta(\mathbf{x}_t, t), \Sigma_\theta(\mathbf{x}_t, t)))
\end{eqnarray}
This allows the model $p_{\theta}$ to only have to estimate the two parameters $\mu$ and $\sigma$ of the estimated denoising step. 
Commonly, $\sigma_{\theta}$ is fixed to a constant variance schedule and is not learnable. This means that in order to estimate $p_{\theta}$, a model needs to learn $\mu_{\theta}(x_t, t)$. 
Electrical wave dynamics can be treated as image-like data and the U-Net architecture from Dhariwal and Nichol \cite{Dhariwal2021} is used to estimate the noise at each step of the reverse diffusion process. 
The model is trained using pairs taken from the forward diffusion process $x_t$, $x_{t-1}$ and taking the mean squared error (MSE) between the noise estimated by the model and the true noise at that step. 

We implemented 6 different diffusion models for different tasks (Task 1-6), {see results sections \ref{sec:results:parameterspecific}-\ref{sec:results:transients} for additional task-specific details}. 
{Throughout this paper, we refer to conditioned and unconditioned diffusion models. 
Conditioned diffusion models exert a task that is constrained. 
For instance, the parameter-specific model in section \ref{sec:results:parameterspecific} is conditioned by the input parameters which guide the diffusion process to produce certain types of wave pattern, and the inpainting model in section \ref{sec:results:2Dinpainting} is conditioned by the surrounding wave pattern as it needs to fill in the missing parts while matching the pattern to the surrounding pattern. 
Whereas conditioned diffusion models generate wave patterns under certain constraints, an unconditional model can dream up any wave pattern without any guidance.
All diffusion models preprocessed the images by downsampling them to $64\times 64$ in order to reduce memory consumption during training, and then upsampling generated images to $128\times 128$. We did not see a decrease in performance from doing this down- and upsampling.}
The conditioned diffusion models in sections \ref{sec:results:parameterspecific}, \ref{sec:results:evolve}, and \ref{sec:results:2Dinpainting} were implemented following Saharia et al. \cite{Saharia2021} using an implementation by Liangwei Jiang and Yury Belousov \cite{Jiang2022}.
The unconditioned diffusion model in sections \ref{sec:results:similarity} and \ref{sec:results:transients} was implemented following Ho et al. \cite{Ho2020} using the Diffusers library \cite{Diffusers}.
The diffusion model in section \ref{sec:results:3Dheart} was implemented following Zhou et al. \cite{Zhou2021} using the official codebase.
All diffusion models include a U-Net \cite{Ronneberger2015} architecture and were implemented in PyTorch \cite{Pytorch}.

\begin{figure*}[htb]
  \centering
  \includegraphics[width=0.98\textwidth]{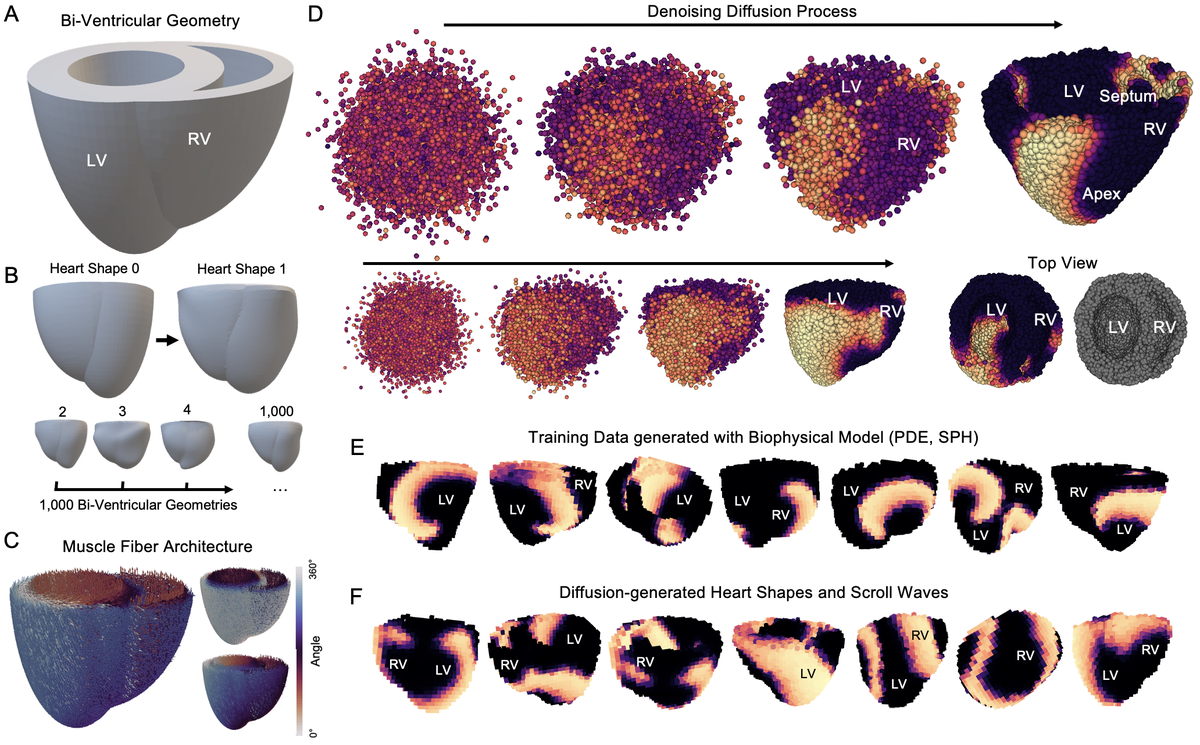}
  \caption{
Diffusion-based modeling of reentrant electrical waves in heart-shaped bi-ventricular geometries (Task 2), see also Supplementary Video 6.
{\textbf{A} Template geometry used in simulations to generate training data.
\textbf{B} $1,000$ randomized, unique variations of template geometry to create unique training samples, as described in \cite{Lebert2023}.
\textbf{C} Geometry-dependent bi-ventricular muscle fiber architecture initiated in each simulation.
\textbf{D} Denoising process during diffusion-based generation of electrical scroll wave pattern in bi-ventricular heart shape. Both the tissue geometry and wave pattern are generated simultaneously.
\textbf{E} Training data used to train diffusion model consisting of $5,000$ training samples showing electrical scroll wave patterns. 
Each simulation consists of $32,000$ particles, subsampled to $16,000$ particles for training, here voxelized and volume-rendered for visualization.
The training data was simulated using a biophysical model (Aliev-Panfilov), see eqs.(\ref{eq:modelu})-(\ref{eq:modelr}), and integrated using the SPH-method \cite{Zhang2021,Zhang2021b}, see section \ref{sec:methods:simulations}. 
\textbf{F} Additional examples of diffusion-generated electrical scroll waves in bi-ventricular heart shapes.
The diffusion model generates a bi-ventricular shape (each shape different) as well as a full dynamical state with both dynamic variables $(u,r)(\vec{x})$. The scroll wave patterns are anisotropic due to the specific ventricular muscle fiber organization in the training data.}
  }
  \label{fig:3Dheart}
\end{figure*}

\begin{table}[htb]
  \begin{tabular}{@{}lrc@{}}\toprule
    Model                & Trainable Parameters        & Training Time      \\\midrule
    {Task 1}             & {$308{,}672{,}266$} & {$2\,\text{days}$} \\
    Task 2             & $31{,}092{,}676$ & $0.5\,\text{day}$ \\
    Task 3             & $62{,}644{,}805$ & $1.5\,\text{days}$ \\
    Task 4      & $965{,}266{,}792$   & $9\,\text{days}$   \\
    Task 5               & $62{,}640{,}193$   & $1.5\,\text{days}$   \\
    Task 6            & $113{,}673{,}219$ & $1\,\text{days}$ \\
    Classification & $11{,}689{,}512$       & $5\,\text{min}$    \\\bottomrule
  \end{tabular}
  \caption{Different model sizes (trainable parameters) and training times used in this study. Training was performed on a single NVIDIA RTX A5000 GPU.}
  \label{tab:training_duration}
\end{table}

\subsection{General Training Details}\label{sec:methods:training}
The networks were trained using the Adam \cite{Adam} optimizer with a learning rate of $10^{-4}$ for the bulk prediction task and $10^{-3}$ for all other tasks. 
We used a batch size of $8$ for the bulk prediction tasks and a batch size of $32$ for all the other tasks.
All neural network models were implemented in PyTorch \cite{Pytorch}.
Training and reconstructions were performed on a NVIDIA RTX A5000 graphics processing unit (GPU), see Table \ref{tab:training_duration} for an overview of training durations.

\subsection{Evaluation}
We evaluated the diffusion models accuracies using the root mean squared error (RMSE), the mean absolute error (MAE), or the multi-resolution perceptual error (MR) \cite{Kazmierczak2022} depending on the model and task.
We computed the errors per frame, averaging over all frames of a separate evaluation dataset that was not part of the training dataset.
While RMSE and MAE correspond to a measure of the average difference per pixel, MR is a measure for the similarity of two patterns and, in more general terms, for how the waves perceptually look to the human eye.
{The issue with RMSE and MAE is that they can produce high errors when images are similar but not perfectly congruent (e.g. a shifted or slightly wider spiral wave pattern, which is otherwise identical).
By contrast, the MR calculates the difference between two images over their embedding in feature space and, therefore, provides a much more holistic comparison of two images over multiple spatial scales and feature hierarchies\cite{Kazmierczak2022}, see also Fig.~\ref{fig:hallucination} and section \ref{sec:results:2Dinpainting}.
We used MR in addition to RMSE and MAE to overcome their limitations related to a crude pixel-wise comparison.
MR captures when two patterns are qualitatively very similar, which RMSE and MAE do not capture per se.}

\section{Results}
\label{sec:results}

\subsection{Parameter-specific Generation of Spiral Wave Dynamics}
\label{sec:results:parameterspecific}

Diffusion models can generate parameter-specific spiral wave patterns when given a set of parameters as input, see Fig.~\ref{fig:parameter-specific1}.
We trained a diffusion model to generate a parameter-specific spatio-temporal spiral wave pattern from $2$ input parameters $D$ and $\epsilon_0$ (Task 1), see also Fig.~\ref{fig:parameter-specific1}D,E):
\begin{eqnarray}
  \label{eq:parameterPrediction}
(\xi(x,y), D, \epsilon_0)  \rightarrow \left( \tilde{u},\tilde{r} \right) \left(x,y, t_1, ... , t_n \right)
\end{eqnarray}
Here, {$\xi(x,y)$ is the initial noise, $D$} and $\epsilon_0$ are parameters of the biophysical Aliev-Panfilov model in eqs. (\ref{eq:modelu})-(\ref{eq:modelr}) which influence the spiral wave properties, and $(\tilde{u},\tilde{r})$ is the generated spatio-temporal spiral wave pattern consisting of multiple timesteps $t_n$ as shown in Fig.~\ref{fig:parameter-specific1}D-F).
The diffusion model generates both dynamic variables $u,r$ and multiple frames ($t_1, ... , t_n$) at once (here usually $n = 5$).
We refer to this generation process as multi-timestep generation 'conditioned by the parameters {$D$} and $\epsilon_0$'.
We found that the generative process can be conditioned and consequently guided by parameters to produce spiral wave dynamics with specific properties, which equally arise with a corresponding biophysical model with the same parameters, see Fig.~\ref{fig:parameter-specific1}B,C). 
Spiral wave shapes can be very different depending on the model parameters, see Fig.~\ref{fig:parameter-specific1}A,B) and Fig.~9 in Qu et al. \cite{Qu2000a} and Figs.~5-9 in Bartocci et al.\cite{Bartocci2011}.
Here, the Aliev-Panfilov model produces spiral waves with wider/thinner arms and longer/shorter diastolic intervals when varying the parameters $D$ and $\epsilon_0$ in eqs.~(\ref{eq:modelu})-(\ref{eq:modelr}), see Fig.~\ref{fig:parameter-specific1}A).
Our diffusion model reproduces these different parameter-specific regimes when conditioned with the respective parameters, see Figs.~\ref{fig:parameter-specific1} and \ref{fig:parameter-specific2}.
Importantly, the diffusion model can generate parameter-specific spiral wave dynamics, even if the parameter combination was not part of the training data.
{However, it fails to generate plausible wave patterns outside of the distribution of training data, see Supplementary Fig. 3, which is generally known to be true for many deep learning models.}

{We performed {$12,500$} unique simulations in total for {$5 \times 5 = 25$} different parameter pairs $(D, \epsilon_0)$ or $500$ simulations per parameter pair.
Simulations were performed with combinations of $(D,\epsilon_0)$ as shown in Fig.~\ref{fig:parameter-specific1}A).
In each simulation, we initialized the spiral waves shown in Fig.~\ref{fig:parameter-specific1}A), and then applied a random number of pacing stimuli (between 10 and 40) in random locations to cause wave break and create single- or multi-spiral wave dynamics as shown in Fig.~\ref{fig:parameter-specific1}B).
Only every $10^{th}$ simulation time step over a period of $10,000$ simulation time steps was written out resulting in $1,000$ frames showing about $2-3$ spiral wave rotations.
Per simulation, we extracted $5$ multi-timestep training samples showing each a unique spatio-temporal spiral wave pattern, yielding $2,500$ samples in total per parameter pair or $62,500$ samples in total over the grid of $5 \times 5$ parameter pairs.
Each sample consists of $n=5$ frames $\{ \{  u (x,y,t_1), r (x,y,t_1) \}, \ldots ,  \{  u (x,y,t_n), r (x,y,t_n) \} \}$ which are $3$ frames apart, effectively covering a period of $t_s = 150$ simulation time steps.
Each sample shows a unique spiral wave pattern, and there is no overlap between training samples.
We augmented the data by randomly flipping or rotating all frames in a sample by multiples of $90^{\circ}$, effectively increasing the training dataset size by a factor of $8$ and ensuring rotational invariance.
Note that most spirals in Fig.~\ref{fig:parameter-specific1}B) are clock-wise rotating as they are simulated data before augmentation.
In principle, the number of frames in a training sample can vary (e.g. $5$, $10$, $15$), but we resorted to $5$ for simplicity.}
{The parameters were first encoded using sinusoidal embeddings\cite{AttentionIsAllYouNeed}. We then conditioned the diffusion model by concatenating these sinusoidal parameter encodings to the diffusion timestep embedding that is passed into each residual connection in the underlying U-Net\cite{Dhariwal2021}, see Fig.~\ref{fig:Fig1}.}
Aside from the parameter conditioning, the generation was unconditioned allowing the diffusion model to dream up any spiral wave pattern.

We verified the parameter-specificity of the diffusion model by initializing the biophysical model from eqs.  (\ref{eq:modelu})-(\ref{eq:modelr}) with the first diffusion-generated frame $\{ \tilde{u} (x,y,t_1), \tilde{r}(x,y,t_1) \}$, see Fig.~\ref{fig:parameter-specific2}A), and integrating the biophysical model for {$t_s=150$ time steps $t_1 \rightarrow t_2 \rightarrow t_3 \rightarrow ... \rightarrow t_{s}$} using either the same parameter combination $(D^*,\epsilon_0^*)$ or a mismatching parameter combination $(D,\epsilon_0)$ to see if the PDE-evolved solutions co-evolve with the spatio-temporal spiral wave pattern generated by the diffusion model, see Fig.~\ref{fig:parameter-specific2}B).
{Note that the diffusion sample times $\{ t_n= 1,2,3,4,5 \}$ correspond to $\{ t_s = 1, 31, 61, 91, 121, 151 \}$ in simulation time steps because of the subsampling during the training data creation (every $10^{th}$ frame stored from simulation) and training procedure (every $3^{rd}$ frame used for one training sample).
We can compare the two solutions because the diffusion-generated spatio-temporal spiral wave pattern shows a plausible spatio-temporal progression of the wave pattern.}
Comparing the simulated state $\{ u (x,y,t_s), r (x,y,t_s) \}$ at time $t_s=151$ to the corresponding {last state $\{ u (x,y,t_n), r (x,y,t_n) \}$} with $t_n=5$ in the diffusion-generated sample, see Fig.~\ref{fig:parameter-specific2}D), we found that the average pixel-wise error (MAE) is smallest with matching parameters $(D,\epsilon_0) = (D^*,\epsilon_0^*)$, see Fig.~\ref{fig:parameter-specific2}E), regardless of whether they were part of the training data or not.
In other words, the diffusion model generates spiral wave dynamics that the biophysical model also produces with the same parameters.
{More precisely, we found that the error was minimal with matching parameters for only $21$ out of the $5 \times 5 = 25$ parameter combinations. 
In the other $4$ cases, it was a nearby combination and the difference in the error was very small.
The $4$ ambiguous cases occurred in the central lower left area of the $5 \times 5$ parameter grid with medium to thin waves. 
The ambiguity could result from the waves being more similar to each other, or, vice versa, harder to distinguish when comparing the divergence of the wave patterns in our measurements. 
Correspondingly, we believe the issue will resolve when including longer trajectories or integration times in our measurements.
The plots in Fig.~\ref{fig:parameter-specific2}E) were derived from averaging over $100$ simulations initialized with the first frames of $100$ different diffusion-generated samples per parameter combination.}
Note that the diffusion model was not trained on all parameter combinations.
The findings suggest that diffusion models do not need to be trained meticulously on all possible parameter combinations, can interpolate in parameter space and generate wave dynamics for many more parameter combinations than just the ones they were trained on.

\subsection{Generation of Reentrant Scroll Waves in Heart-shaped Geometries}
\label{sec:results:3Dheart}

\begin{figure}[htb]
  \centering
  \includegraphics[width=0.48\textwidth]{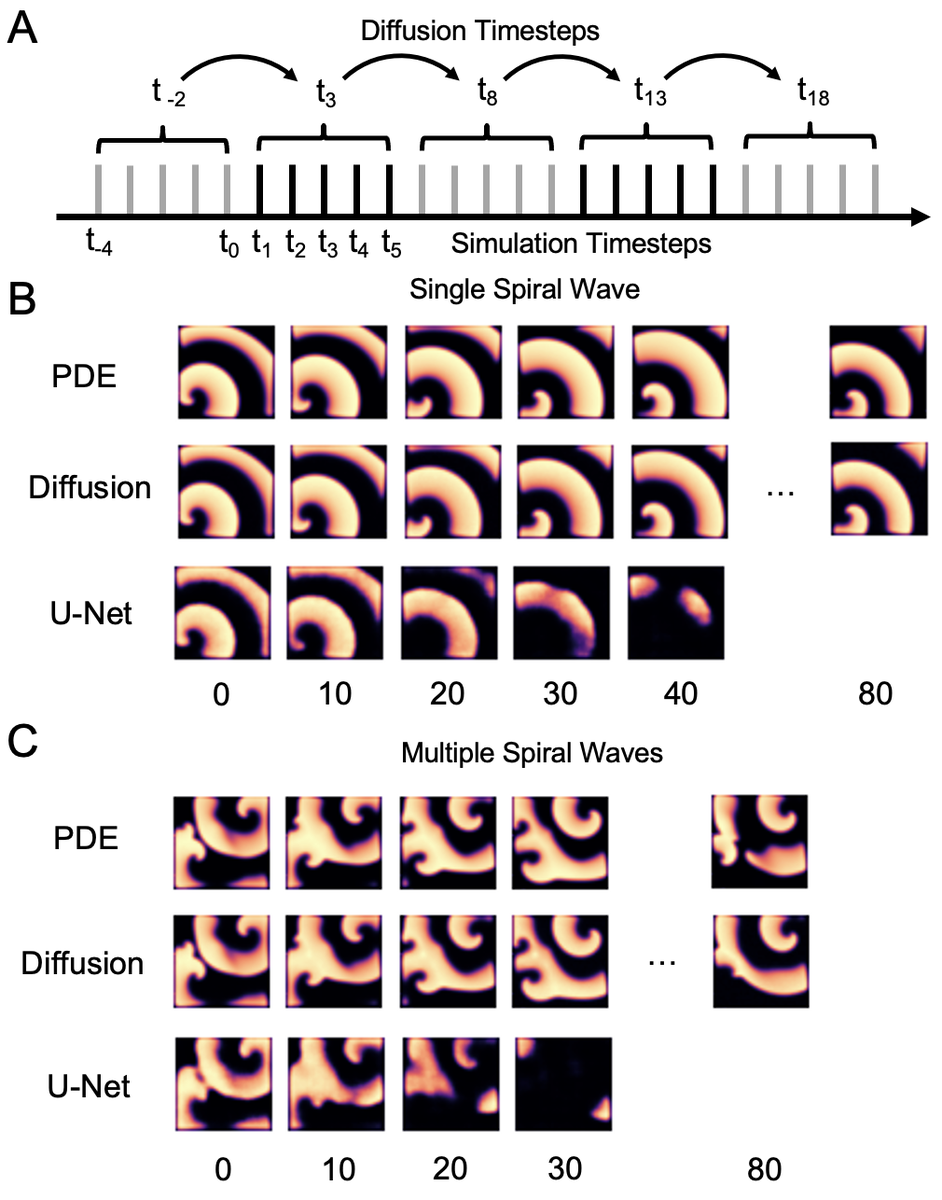}
  \caption{
Data-driven modelling of spiral wave dynamics using diffusion models.
\textbf{A} Spatio-temporal prediction (Task 3) of $5$ frames from previous $5$ frames, see section \ref{sec:results:evolve}.
\textbf{B,C} Comparison of ground-truth data (GT) simulated with biophysical model (finite differences) and data-driven methods (Diffusion vs. U-Net) to evolve the wave pattern.
With a single spiral wave, the output of the diffusion model is visually indistinguishable from the ground-truth for many rotations, while U-Net quickly fails to sustain the wave pattern.
With more complicated wave patterns, the diffusion models begins to deviate from the biophysical model after 2-3 spiral rotations (80 time steps).
  }
  \label{fig:evolution}
\end{figure}

We trained a diffusion model to generate scroll waves in bi-ventricular-shaped geometries (Task 2), as shown in Fig.~\ref{fig:3Dheart}.
{Panel D) shows the denoising diffusion process used to generate bi-ventricular-shaped point clouds with corresponding excitation values per point, where both the shape and the electrical wave pattern are generated simultaneously by the diffusion model (two representative examples).}
Fig.~\ref{fig:3Dheart}F) shows further examples of diffusion-generated scroll waves, which are visually indistinguishable from the scroll wave patterns shown in E) which were simulated using the biophysical model in eqs. (\ref{eq:modelu}-\ref{eq:modelr}).
The generative process was completely unconstrained and not explicitly conditioned.
Correspondingly, the model generates any scroll wave pattern in any bi-ventricular shape it can come up with given what it has learned from the training data.
The scroll wave patterns are anisotropic because the training data was simulated with an anisotropic ventricular fiber architecture.
The training data, therefore, implicitly warrants anisotropy during the generations.
Even though the diffusion model generates only a single scroll wave pattern $u(\vec{x},t)$, {and only the excitatory variable $u$}, it is easy to imagine how this pattern could also be evolved over time  $u(\vec{x},t_1,t_2, \ldots)$ as described in the next section \ref{sec:results:evolve} and shown in Fig.~\ref{fig:evolution}.

{Simulations were performed as described in Lebert et al.\cite{Lebert2023}.
Accordingly, the simulated training data consists of point clouds of $i \sim 32,000$ vertices $p(\vec{x})_i$ representing bi-ventricular heart shapes. 
The simulations were performed with $1,000$ unique bi-ventricular shapes created from a template geometry, see Fig.~\ref{fig:3Dheart}A,B).
Accordingly, the diffusion model comes up with similar shapes during the generative process.
The excitatory variable $u_i$ is defined per vertex $i$.
We downsampled the data to $16,000$ points and used Point-Voxel Diffusion \cite{Zhou2021} trained on $5,000$ training samples obtained from the simulations where each training sample consists of a single point cloud of excitation values $u (\vec{x,t})$ at a particular time $t$.
We trained the model to output $16,384$ points (with a latent dimension of $512$) for $400$ epochs.}

\subsection{Generative Diffusion-based Simulation of Electrical Wave Dynamics}
\label{sec:results:evolve}

Electrical impulse propagation in the heart is usually simulated by integrating partial differential equations in space and over time by using, for example, the finite difference or finite element methods.
We trained a diffusion model to calculate an immediate future time step of a given spatio-temporal excitation wave pattern (Task 3):
 \begin{eqnarray}
  \label{eq:2Devolution}
(u,r)(\vec{x},t) \rightarrow (\tilde{u},\tilde{r})(\vec{x},t+\tau)
\end{eqnarray}
where $(u,r)$ are the dynamic variables from eqs.~(\ref{eq:modelu}-\ref{eq:modelr}) and $\tau$ is an infinitesimal temporal increment or the integration time step.
In other words, we employed diffusion-based data-driven modeling to evolve electrical wave dynamics over time rather than using a biophysical model to simulate the dynamics. 
More precisely, we trained a diffusion model to predict {the next $5$ time steps from the previous $5$ time steps of the dynamics}, resulting in an integration scheme that updates a brief spatio-temporal pattern instead of a static spatial pattern:
 \begin{eqnarray}
  \label{eq:2Devolutionb}
(u,r)(\vec{x},t_{-4}, \ldots, t_{0}) \rightarrow (\tilde{u},\tilde{r})(\vec{x},t_1 , \ldots , t_5)
\end{eqnarray}
Here, $\{ t_{-4}, t_{-3}, t_{-2} , t_{-1}, t_0 \}$ are the $4$ previous time steps and the current time step $t_0$ and $\{ t_1 , \ldots , t_5 \}$ are the next $5$ time steps predicted by the model, see Fig.~\ref{fig:evolution}A).
We found this multi-timestep prediction scheme more stable than updating the dynamics one time step at a time or predicting the next time step from the previous $n$ time steps in an auto-regressive manner.
{
We found empirically that using $5$ time steps to predict the next $5$ time steps was a good compromise between performance and training time. 
We tested using $10$ time steps to predict the next $10$ time steps, which worked as well (and presumably better), but the model was consequently bigger and the training time much larger. 
In all cases, we used no temporal subsampling. 
We conditioned the diffusion model by concatenating $(u,r)(\vec{x},t_{-4}, \ldots, t_{0})$ to the initial noisy distribution $\xi (x,y)$, adding five channels to the input of the underlying U-Net ($10 \times 128 \times 128$ pixels).}
We trained and evaluated the model with $15,000$ and $5,000$ samples, respectively.
The training samples show either simple or complex two-dimensional spiral wave dynamics simulated with two parameter sets, see Table \ref{tab:parameters} (Task 3a/b) and Fig.~\ref{fig:evolution}B,C). 
The same parameters were also used in Task 5 and in Fig.~\ref{fig:hallucination}.
{For each of the two parameter sets, we ran $100$ simulations and sampled the training samples from $75$ simulations and the evaluation samples from $25$ simulations, respectively.}

Fig.~\ref{fig:evolution}B,C) shows the ground-truth (PDE) spiral wave patterns for up to $80$ simulation time steps and the corresponding evolved spiral wave patterns predicted either with our diffusion model or a correspondingly trained U-Net model.
While the dynamics quickly degenerate with U-Net, the diffusion model successfully sustains and evolves the dynamics over a very long time.
The ability to sustain the wave pattern is likely related to diffusion models being able to learn and mimick shapes.
The diffusion model produces spiral waves with either stable or meandering cores which exhibit breakup and (self-) interactions.
With the single spiral wave in Fig.~\ref{fig:evolution}B) the diffusion model's output matches the biophysical model's output for many rotations, see Supplementary Video 7.
Eventually, the original (ground-truth) dynamics diverge from the diffusion-generated dynamics, which is, to some extent, to be expected as the dynamics would also diverge with, for instance, two different classical integration methods.
With the more complex spiral wave dynamics in Fig.~\ref{fig:evolution}C) the diffusion output diverges rapidly within less than 2 rotations of the spiral wave pattern, see Supplementary Video 8.
{Interestingly, the diffusion model appears to favor more stable wave dynamics (less wave break), see right panel in Fig.~\ref{fig:evolution}C).} 
We can only speculate that this could be related to a bias in the training data, {e.g. an underrepresentation of finer spatial scales}, see also section \ref{sec:results:transients} for further details regarding the ensemble behavior of the dynamics.

{The diffusion-based time stepping appears to only work well with spatio-temporal data, which suggests that spatio-temporal data is unique enough so that the model is sufficiently constrained (or conditioned), which in turn enables it to predict the next spatio-temporal segment reasonably well. 
However, how these findings generalize to various dynamical regimes with different Lyapunov times warrants further research.}

Updating the dynamics in a $128 \times 128$ pixel simulation domain takes $1.1 \, ms$ on a NVIDIA A5000 GPU {per multi-frame prediction}. 
Together with the results in \ref{sec:results:3Dheart}, our findings suggest that diffusion-based modeling could be used to simulate spatio-temporal dynamics in the heart.

\begin{figure*}[htb]
  \centering
  \includegraphics[width=0.98\textwidth]{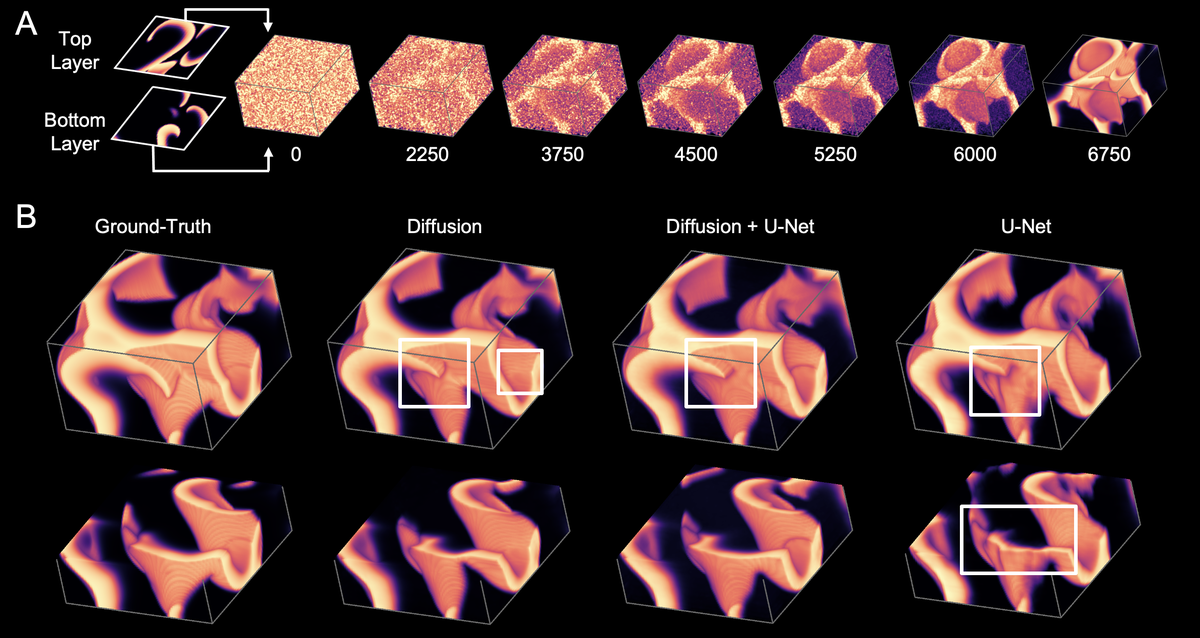}
  \caption{
Diffusion-based reconstruction of scroll wave dynamics inside a three-dimensional bulk from two-dimensional observations of the dynamics on the bulk's top and bottom surface (Task 4), see also Supplementary Video 9. The bulk is fully opaque and measurements can only be obtained from its surface.
\textbf{A} Illustration of diffusion process over denoising iterations.
\textbf{B} Scroll wave dynamics (left: ground-truth) and reconstructed scroll waves with diffusion (center left), U-Net (right) and U-Net refined with diffusion (center right), see also \cite{Lebert2023,Stenger2023}. 
The U-Net is from our previous study\cite{Lebert2023}.
{The top row shows the full bulk while the bottom row shows the lower half of the bulk to highlight the central midwall layer of the bulk, which exhibits the highest reconstruction error.}
While diffusion produces smoother wave patterns than U-Net, particularly at deeper layers, the overall reconstruction accuracies are not significantly different across the three approaches, see also Fig.~\ref{fig:3D2}.
White squares highlight slight differences between reconstructions and ground-truth.
The bulk is $128 \times 128 \times 40$ voxels (aspect ratio was altered to emphasize transmural wave morphology), see also \cite{Lebert2023}. The simulation parameters are shown in Table \ref{tab:parameters} (Task 4).
  }
  \label{fig:3D1}
\end{figure*}

\subsection{Reconstruction of Three-Dimensional Scroll Wave Dynamics from Surface Observations}
\label{sec:results:3D}

Measuring electrophysiological wave phenomena beneath the heart surface is a long-standing challenge in cardiovascular research and diagnostics. 
Catheter electrodes or optical mapping provide only superficial data from the heart surface, and intramural measurements from within the heart muscle with electrodes are sparse.
To address this challenge, various numerical methods were introduced which aim at reconstructing transmural wave patterns from observations of the dynamics on the tissue's surface \cite{Hoffman2016,Hoffman2020,Lebert2023,Stenger2023}.
The numerical reconstructions are particularly relevant in the context of tachyarrhythmias, such as ventricular or atrial fibrillation, as they may provide a better understanding of the underlying three-dimensional spatio-temporal organization of the electrical waves within the heart muscle.
Recently, Lebert et al. \cite{Lebert2023} and Stenger at al. \cite{Stenger2023} demonstrated that convolutional encoding-decoding neural networks (different U-Net-types) can be used to reconstruct three-dimensional scroll wave dynamics inside a thick bulk-shaped excitable medium from two-dimensional observations of the dynamics on the top and/or bottom surfaces (representing the epi- and endocardium). 
At the same time, Stenger et al. demonstrated this briefly also with a diffusion model \cite{Stenger2023}.
However, several aspects of the deep learning-based reconstructions remain underexplored, in particular with the diffusion-based approach.

We trained a diffusion model to predict three-dimensional scroll wave dynamics inside a bulk from two-dimensional observations of the dynamics on the surface of the bulk (Task 4), as described in Lebert et al. \cite{Lebert2023} and shown in Fig.~\ref{fig:3D1}.
More specifically, the model was trained to predict a single three-dimensional snapshot of the excitatory variable $u_t(x,y,z)$ at a given time $t$ at every voxel in a bulk with $128 \times 128 \times 40$ voxels from the $5$ {previous} two-dimensional snapshots of the dynamics on the bulk's surface:
 \begin{eqnarray}
  \label{eq:3Dprediction}
  \left(u_1(x, y), \ldots, u_5(x, y)\right) \rightarrow \tilde{u}_5(x, y, z)
\end{eqnarray}
where $\tilde{u}$ is a prediction of the true dynamics $u$.
The snapshots were measured either i) on the top surface only (single-surface mode) resulting in a spatio-temporal measurement consisting of $5$ snapshots $(u_1(x,y,1), \ldots , u_5(x,y,1))$ or ii) on the top and bottom surface (dual-surface mode) resulting in $2 \cdot 5$ snapshots $(u_1(x, y, 1), u_1(x, y, 40), u_2(x, y, 1), \ldots , u_5(x, y, 40))$, as described in Lebert et al. \cite{Lebert2023}.
The $5$ snapshots were sampled at equidistant times at $t_{-4 \tau}, t_{-3 \tau}, t_{- 2 \tau}, t_{-\tau}, t_0$ with $u_i = u(t_i)$ over about one rotational period $T$ of the scroll wave dynamics ($\tau \approx T/5$), which we found to provide sufficient information to reconstruct the dynamics, as described in \cite{Lebert2023,Christoph2020,Lebert2021}.
Accordingly, we conditioned the diffusion model by concatenating these sequences as additional channels in the U-Net inputs (interleaved in dual-surface mode, odd indices for the top layer and even indices for the bottom layer). 

\begin{figure}[htb]
  \centering
  \includegraphics[width=0.48\textwidth]{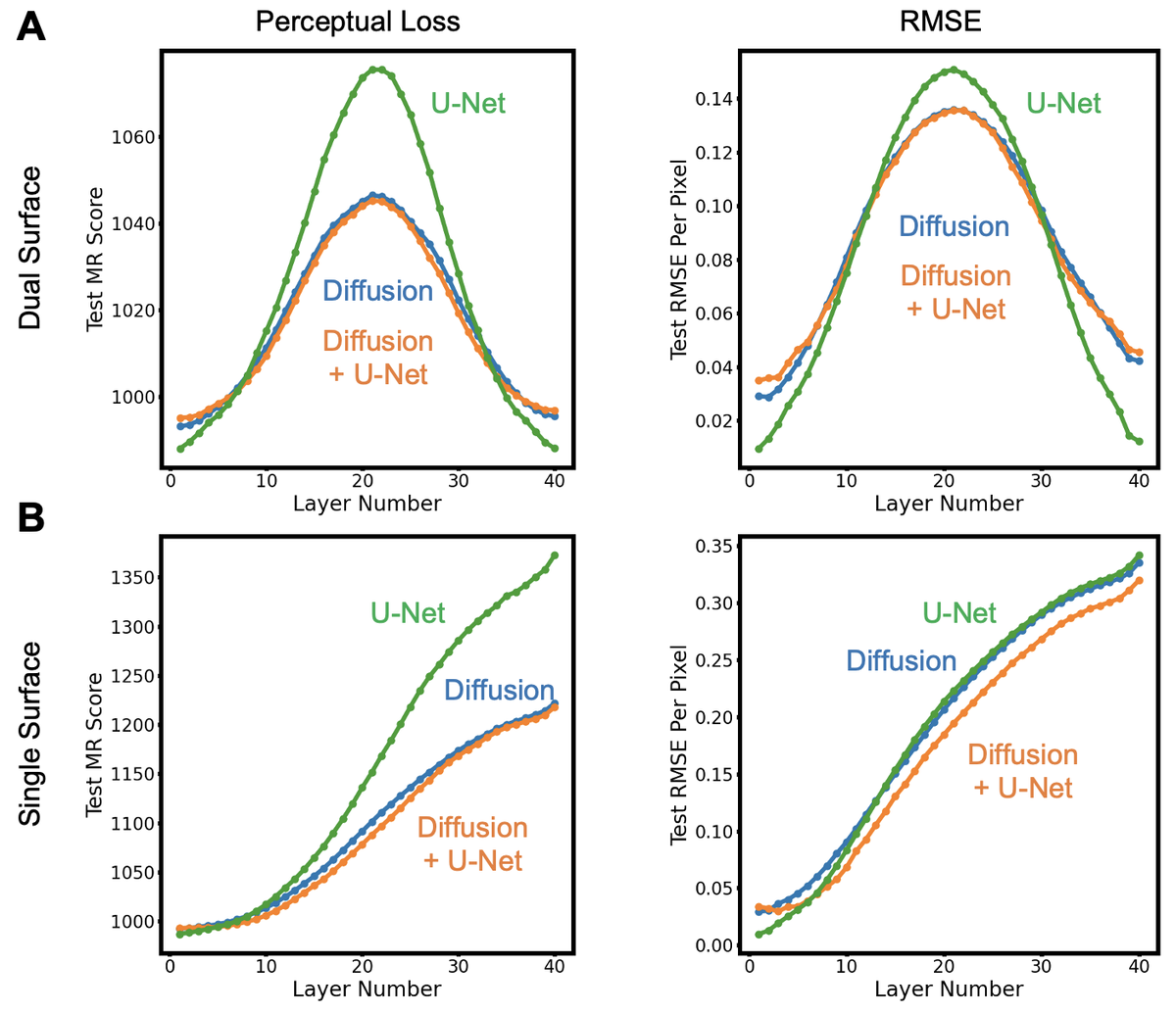}
  \caption{
Transmural reconstruction error per layer number or depth with diffusion (blue), U-Net (green, from \citet{Lebert2023}) or diffusion + U-Net (orange), see also Fig.~\ref{fig:3D1}.
The reconstruction error was quantified using either the perceptual error \cite{Kazmierczak2022} (left) or absolute difference (RMSE, right).
\textbf{A} Dual-surface reconstruction analyzing top and bottom layers: U-Net performs worse at midwall, but slightly better closer to the surfaces than diffusion.
Perceptual error and RMSE produce very different error profiles.
\textbf{B} Single-surface prediction: the perceptual error increases more steeply with U-Net as with diffusion, while all models perform similar with RMSE.
  }
  \label{fig:3D2}
\end{figure}

\begin{figure*}[htb]
  \centering
  \includegraphics[width=0.98\textwidth]{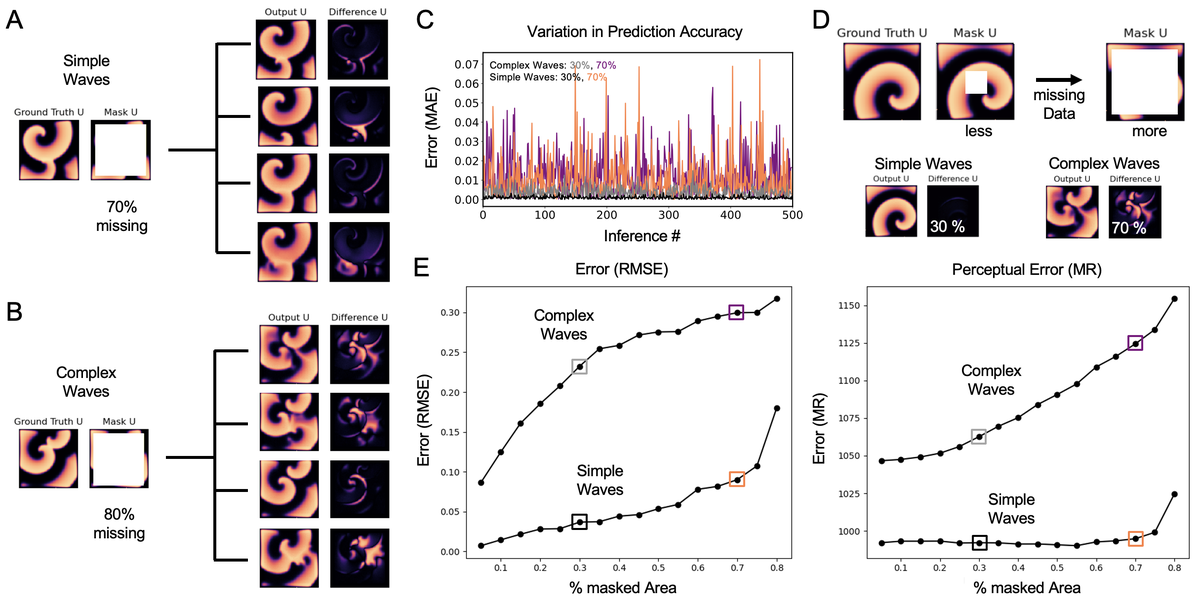}
  \caption{
Hallucination of diffusion model during inpainting of electrical spiral wave dynamics (Task 5). 
\textbf{A} The diffusion model predicts missing data inside a square region (white) at the center of the medium. 
With most of the data missing (70\%), the diffusion model dreams up wave patterns which look convincingly like spiral wave patterns but deviate substantially from the ground-truth: 4 repetitive predictions for the same interpolation task.
\textbf{B} The effect becomes more severe with more complex waves and fewer data (80\% missing), see also panel D and Supplementary Video 10.
\textbf{C} The model generates significantly different output when the same prediction task is performed repeatedly (500 times, black: simple waves with 30\%, gray: simple waves with 30\%, orange: complex waves with 70\%, pink: complex waves with 70\% missing data, respectively).
\textbf{D} Combinations of less vs. more missing data (30\% vs. 70\%) and simple and complex waves.
\textbf{E} Average prediction error (RMSE vs. perceptual MR) with increasing percentage of missing data.
Hallucination is minimal with simpler wave patterns and 10-30\% missing data and increases steeply with complex wave patterns.
The perceptual error indicates a higher {qualitative} agreement between inpainted and ground-truth patterns which is not reflected by RMSE, see also Fig.~\ref{fig:3D2}.
  }
  \label{fig:hallucination}
\end{figure*}

To explore an alternative extension of our reconstruction approach, we conditioned the diffusion model using the output of a generic U-Net model, which was trained and applied as described in \cite{Lebert2023}, to create a combined model that potentially can take advantage of the strengths of both the U-Net and diffusion models, see also Fig.~\ref{fig:3D2}.
Accordingly, we conditioned the combined model with the sequences of $5$ (or $2 \cdot 5$) two-dimensional snapshots and one three-dimensional prediction $\tilde{u}(x,y,z)$ of the U-Net model, which analyzed in turn also $5$ snapshots as input. The two- and three-dimensional inputs were concatenated to obtain $(128 \times 128 \times 45)$ or ($128 \times 128 \times 50$) samples in single- vs. dual-surface mode as conditions, respectively.
This leads to a total of 4 conditioning modes that we tested (single- vs. dual-surface, diffusion vs. combination of diffusion + U-Net).
Generally, the different model versions required three-dimensional input samples, e.g. $(128 \times 128 \times 5)$ or $(128 \times 128 \times 45)$.
To denoise a $128 \times 128 \times 40$ volume image with $5 \cdot 128 \times 128$ snapshots as conditioning, the model corresponds to an $R(128 \times 128 \times 45) \rightarrow R(128 \times 128 \times 40)$ function. 
However, internally, because the denoising diffusion process works on the intermediate noisy bulk data, the overall data consists of the conditioning data concatenated to the noisy data, which then results in an array size of, for example, $128 \times 128 \times 80$.

Training was performed with $20,000$ training samples, which were generated in $100$ simulations, see also section \ref{sec:methods:simulations}, and the model was evaluated on $5,000$ separate samples.
The simulated bulk is completely opaque (only surface voxels can be observed) and thick enough to sustain three-dimensional scroll wave dynamics ($128 \times 128 \times 40$ voxels, 1-2 scroll wavelengths), see Fig.~\ref{fig:3D1}.
We used the same simulation data and parameters as in Lebert et al.\cite{Lebert2023}, see also Table \ref{tab:parameters} (Task 4).

Fig.~\ref{fig:3D1}A) shows the denoising process during the scroll wave prediction task in the bulk using the diffusion model.
The scroll wave pattern is reconstructed from the top and bottom surface layers.
Interestingly, the rough shape of the scroll wave pattern is already captured early in the denoising process, while later stages enhance finer structures.
Fig.~\ref{fig:3D1}B) shows a comparison of the predictions obtained with diffusion vs. U-Net vs. the combination of the two with diffusion refining the U-Net output, see also Fig.~\ref{fig:3D2}.
All models are able to predict three-dimensional scroll waves from two-dimensional observations using either only the top (single-surface mode) or both the top and bottom surface layers (dual-surface mode), see also Fig.~\ref{fig:3D2}.
The diffusion model slightly outperforms U-Net, but their combination does not significantly increase the reconstruction accuracy beyond the accuracy of the diffusion model.
In \citet{Stenger2023} it was found that diffusion performs substantially better than U-Net with long observations ($32$ snapshots).
Here, we used fewer observations (only $5$ snapshots), which likely causes this discrepancy.

Most importantly, the diffusion-based reconstructions exhibit one striking feature: while U-Net reconstructions become fuzzier with increasing depth, diffusion maintains the shape, smoothness, and overall look of the scroll waves much better throughout the bulk.
This is also reflected by the perceptual error, see Fig.~\ref{fig:3D2} and also section \ref{sec:results:2Dinpainting}.
However, even though the visual impression suggests otherwise, we find, on average, no dramatic improvement of the overall reconstruction accuracy (RMSE) with diffusion over U-Net, see Fig.~\ref{fig:3D2}.
Upon closer inspection, one notices that diffusion produces minor mismatches at deeper layers (white boxes in Fig.~\ref{fig:3D1}B), suggesting that its output looks better than it is and is not necessarily more accurate than with U-Net, see also section \ref{sec:results:2Dinpainting}.
We hoped that guiding the diffusion model with the output from the U-Net model could mitigate these issues, but, on average, the error remained the same, see Fig.~\ref{fig:3D2}.
Unlike in \citet{Stenger2023}, our model produces smooth scroll wave patterns without residual noise.
Our diffusion model predicts the bulk at once and not layer by layer, which could cause the smoother appearance of the waves.

It is important to highlight that our deep learning-based scroll wave reconstruction approach was only trained on the Aliev-Panfilov scroll wave dynamics and, therefore, assumes scroll waves inside the tissue. 
All three types of neural networks were trained with thousands of corresponding pairs of three- and two-dimensional data of scroll waves and observations thereof.
Therefore, the training data implicitly restricts the approach to a particular distribution of data and its characteristics (specific electrophysiological model that produces waves with a particular shape, isotropic vs. anisotropic wave patterns, wavelength relative to medium thickness), and the approach is task-specific (single- vs. dual-surface observations).
{It would be interesting to test how the reconstructions perform and what type of waves they produce with significantly different data.}

\subsection{Hallucination during Inpainting of 2D Spiral Wave Dynamics}
\label{sec:results:2Dinpainting}
Generative models are known to hallucinate, which means that they may generate output that looks, sounds or reads convincing but is inaccurate \cite{Cohen2024, Kalai2024}. 
For example, recent large language models (LLMs) are known to confidently present made-up knowledge as if it was factual.
In computer vision, diffusion models may generate unexpected scenes which are abstract and not part of human day-to-day experience. 
While it is easy to identify hallucination in diffusion-generated visual scenes, it is not necessarily obvious with spiral or scroll wave patterns when they include hallucinations, see also section \ref{sec:results:similarity}.
In Fig.~\ref{fig:3D1}, the diffusion-generated reconstructed scroll wave patterns at midwall look convincing and can be misinterpreted as accurate solutions, but they are just as inaccurate as the output from the U-Net model.

We explored this hallucinating behavior further in a two-dimensional inpainting task of spiral wave dynamics (Task 5), see Fig.~\ref{fig:hallucination}, and can confirm that hallucination occurs, particularly when the task is insufficiently constrained, see also Supplementary Video 10.
We varied the size of a square region at the center of the medium, within which the diffusion model was tasked to interpolate the missing spiral wave pattern from the surrounding data.
Hallucination is minimal with a small square, which is reflected by the error (RMSE) on the left sides of the graphs in Fig.~\ref{fig:hallucination}E). 
However, we observed that the diffusion model comes up with many different spiral wave patterns when the square region is large or more data is missing, see Fig.~\ref{fig:hallucination}A,B) and Supplementary Video 10.
Fig.~\ref{fig:hallucination}C) shows the variation of the diffusion model output when the same task is repeated 500 times with simple vs. complex waves and 30\% vs. 70\% missing data, respectively.
The variation in the output as quantified by the error (MAE) between the ground-truth and the individual predicted spiral wave pattern is particularly large with 70\% missing data.
Hallucination becomes also stronger when the wave dynamics are more complicated (we tested two parameter sets, see Table \ref{tab:parameters}, Task5a/b), compare panels A,B) and the upper and lower curves in E) (average error calculated over $500$ unique samples per data point).
In Fig.~\ref{fig:hallucination}E) we compared the root mean squared error (RMSE), which reflects the pixel-wise congruency of the ground-truth and the predicted pattern, with a perceptual error (MR), which reflects similarities or differences in the patterns independently from spatial mismatches (as it is calculated on the embedding of the pattern).
The perceptual error indicates that with simple waves the variations in the output of the diffusion model are small regardless of the size of the masked area.
In other words, differences in the waves only correspond to slight spatial mismatches (which cause high RMSE), while the wave shapes are very similar qualitatively.
The spikes in C) (with 70\%), on the other hand, correspond to large qualitative changes of the wave pattern which occur occasionally with both complex and simple waves.
Our data suggests that i) the diffusion model hallucinates if it has the freedom to generate many potential solutions to a problem, ii) hallucination can be associated with qualitative changes in the topology of the wave pattern, and iii) hallucination can be mitigated by sufficiently constraining the task the diffusion model is supposed to perform.

The inpainting diffusion model (Task 5) was trained as follows: 
we masked or left out data from a square region at the center of the $128 \times 128$ pixel simulation domain and trained the network with corresponding pairs of masked $u_m(x,y)$ and ground-truth data $u(x,y)$ to reconstruct the missing parts of the spiral wave pattern:
 \begin{eqnarray}
  \label{eq:2Dinpainting}
\left( u_{m,1}(x,y), \ldots, u_{m,5}(x,y) \right) \rightarrow \tilde{u}_5(x, y)
\end{eqnarray}
where $\tilde{u}$ is a prediction of the ground-truth dynamics $u$ and the model reads a short spatio-temporal sequence of $5$ two-dimensional snapshots.
Masked pixels were replaced by zeros. {We conditioned the diffusion model by concatenating $\left( u_{m,1}(x,y), \ldots, u_{m,5}(x,y) \right)$ to the initial noisy distribution $\xi (x,y)$, adding five channels to the input of the underlying U-Net ($6 \times 128 \times 128$ pixels).}
We simulated two spiral wave regimes and trained two separate models: i) one with largely only one spiral wave (Task 5a) and ii) one with multiple, more chaotic spiral waves (Task 5b), see Table \ref{tab:parameters} for the corresponding simulation parameters.
Both models were trained equally with a range of masks with uniform distribution (of the percentage of masked area vs. total area).
We varied mask sizes $m \in [0.05, \ldots, 0.8]$ (percentage masked area vs. total area from 5\% to 80\% in increments of 5\%).
Training and evaluation was performed with $27,500$ and $6,000$ samples, respectively.
{The training samples were drawn randomly from $100$ simulations performed with the biophysical model defined in eqs. (1-2) for each regime, and the evaluation samples were drawn from $25$ separate simulations for each regime.}

\begin{figure}[htb]
  \centering
  \includegraphics[width=0.48\textwidth]{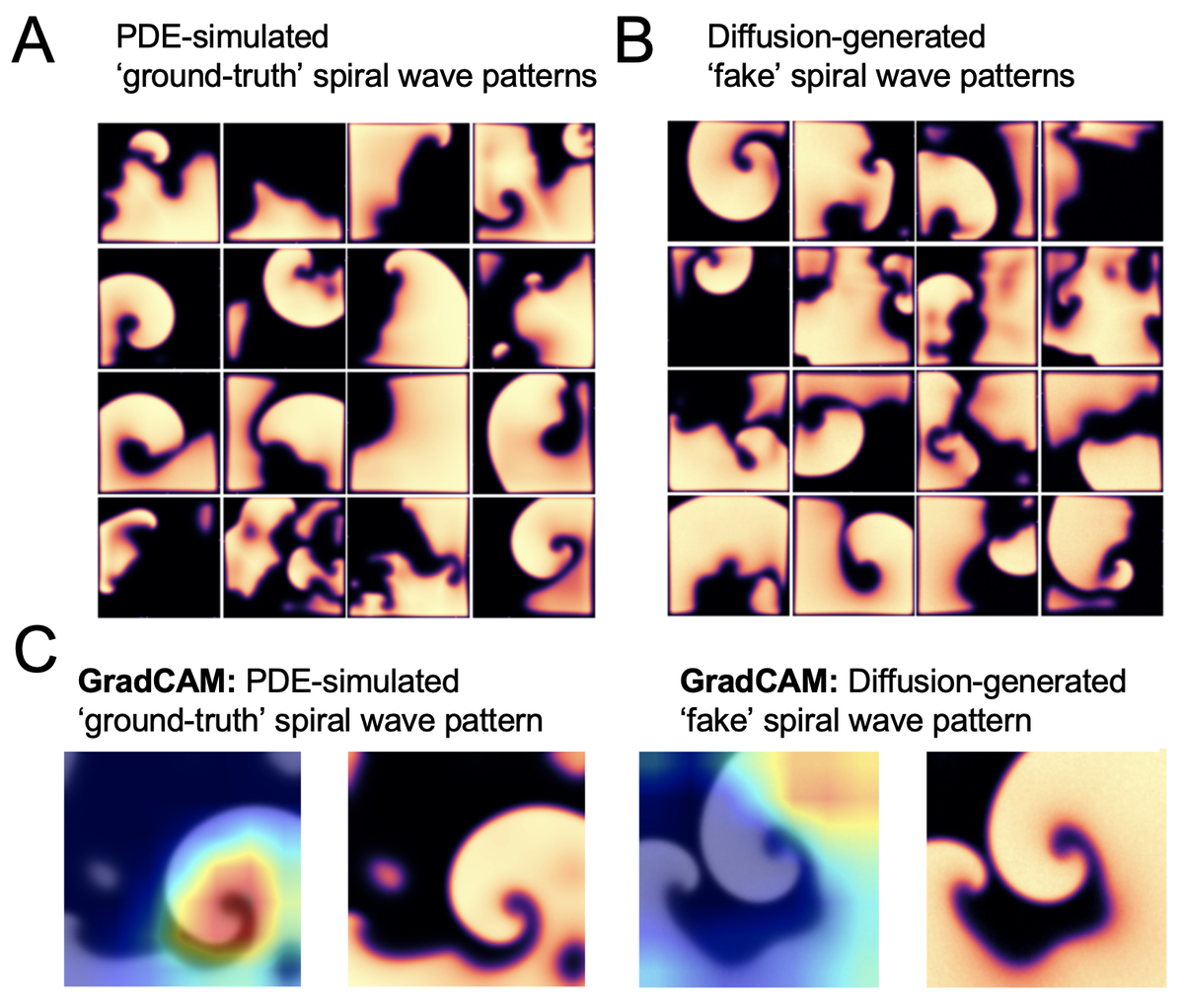}
  \caption{
Comparison of \textbf{A} spiral wave patterns simulated in computer simulations using a biophysical model (PDE, 'ground-truth') and \textbf{B} 'fake' spiral wave patterns generated with diffusion model from eq.~(\ref{eq:2Ddreaming}) (unconditional generation, Task 6). 
The two datasets are visually indistinguishable from each other.
\textbf{C} {ResNet18 \cite{He2015} classifier can distinguish real from 'fake' spiral waves with $99\%$ accuracy. GradCAM\cite{Selvaraju2019} highlights features to which the classification is sensitive.}
  }
  \label{fig:2Dclassification}
\end{figure}

\subsection{Visual Similarity of Diffusion- vs. PDE-generated Spiral Waves}
\label{sec:results:similarity}
Diffusion-generated 'fake' spiral waves are visually indistinguishable from real spiral wave patterns simulated with a biophysical model, see Figs.~\ref{fig:parameter-specific1}B,C) and \ref{fig:2Dclassification}A,B).
Each panel in Figs.~\ref{fig:parameter-specific1}B,C) and \ref{fig:2Dclassification}A,B) shows randomly chosen, representative examples of spiral wave patterns simulated with a biophysical model or generated with diffusion, respectively.
The biophysical model integrates partial differential equations (PDEs), whereas the diffusion model mimics these solutions.
We found it impossible to distinguish diffusion-generated from PDE-simulated spiral wave patterns visually (we tested this systematically with different lab members).
However, despite the visual similarity, a ResNet18 \cite{He2015} classifier fine-tuned on the two classes of images in Fig.~\ref{fig:2Dclassification}A,B) is able to distinguish the two groups of spiral wave patterns with an accuracy of 99.7\% (separate training and validation/test datasets).
This may be due to invisible artifacts from the denoising process or the capability of CNNs to learn minute differences between classes. 
This is well explored\cite{Wang2020, Hou2024}, and was in part the motivation behind the joint generator-discriminator training process of GANs.
{An analysis of the fine-tuned ResNet18 classifier using GradCAM\cite{Selvaraju2019} provides some insights into the classification mechanism but is overall difficult to interpret and remains inconclusive, see Fig.~\ref{fig:2Dclassification}C) and Supplementary Fig.~3.}

{It is only possible to visually distinguish real from 'fake' spiral waves when the model was not trained for long enough.
In that case, the generations often include noisy spiral wave images, see Supplementary Fig.~1.
The training dataset size does not seem to impact the image quality: both models trained with small ($100-1,000$ samples) and large training datasets (more than $10,000$ samples) exhibit noisy images if they are not trained for long enough.
Otherwise, training our diffusion models with sufficiently large training datasets is required to cover a wider range of the many possible, highly chaotic, and diverse spiral wave patterns.
We found that diffusion models can generate plausible-looking spiral wave patterns with as few as $100$ training samples, see Supplementary Fig.~2.
Interestingly, even the parameter-specific model can generate spiral wave patterns when the training samples are distributed over the $25$ parameter pairs ($4$ samples per pair).
Nevertheless, when calculating the "Fréchet Inception Distance" or "FID Score"\cite{Heusel2018}, which measures how visually similar the generated and real images are, and how well the generated images capture the entire distribution of real images, we found that at least tens of thousands of training samples are necessary for training, see Table \ref{tab:training_dataset_size}.}

\begin{table}[htb]
  \begin{tabular}{@{}rr@{}}\toprule
    Training Samples               & FID Score       \\\midrule
    $100$             & $27.54$ \\
    $500$             & $12.22$ \\
    $1,000$      & $8.77$   \\
    $5,000$      & $8.34$   \\
    $10,000$      & $7.01$   \\
    $50,000$ & $7.05$  \\\bottomrule
  \end{tabular}
  \caption{Training dataset size evaluated using "Fréchet Inception Distance" or "FID score"\cite{Heusel2018}. The lower the score, the better the similarity between the real and diffusion-generated images measured over the entire distribution of images. Training dataset from Fig.~\ref{fig:2Dclassification}.}
  \label{tab:training_dataset_size}
\end{table}

{Further, we found that the parameter-specific model with which we generated the patterns in Fig.~\ref{fig:parameter-specific1}C) required more training samples and training iterations than the unconditional model with which we generated the patterns in Fig.~\ref{fig:2Dclassification}B).
Further research is needed on the data-efficiency of diffusion models in these different applications. 
Data-efficiency is an important consideration when looking at the possibility of fine-tuning or training diffusion models from scratch on experimental data. 
It is possible that diffusion models require a large amount of data to perform well in complex applications, which could be a major limitation.}

\begin{figure}[htb]
  \centering
  \includegraphics[width=0.48\textwidth]{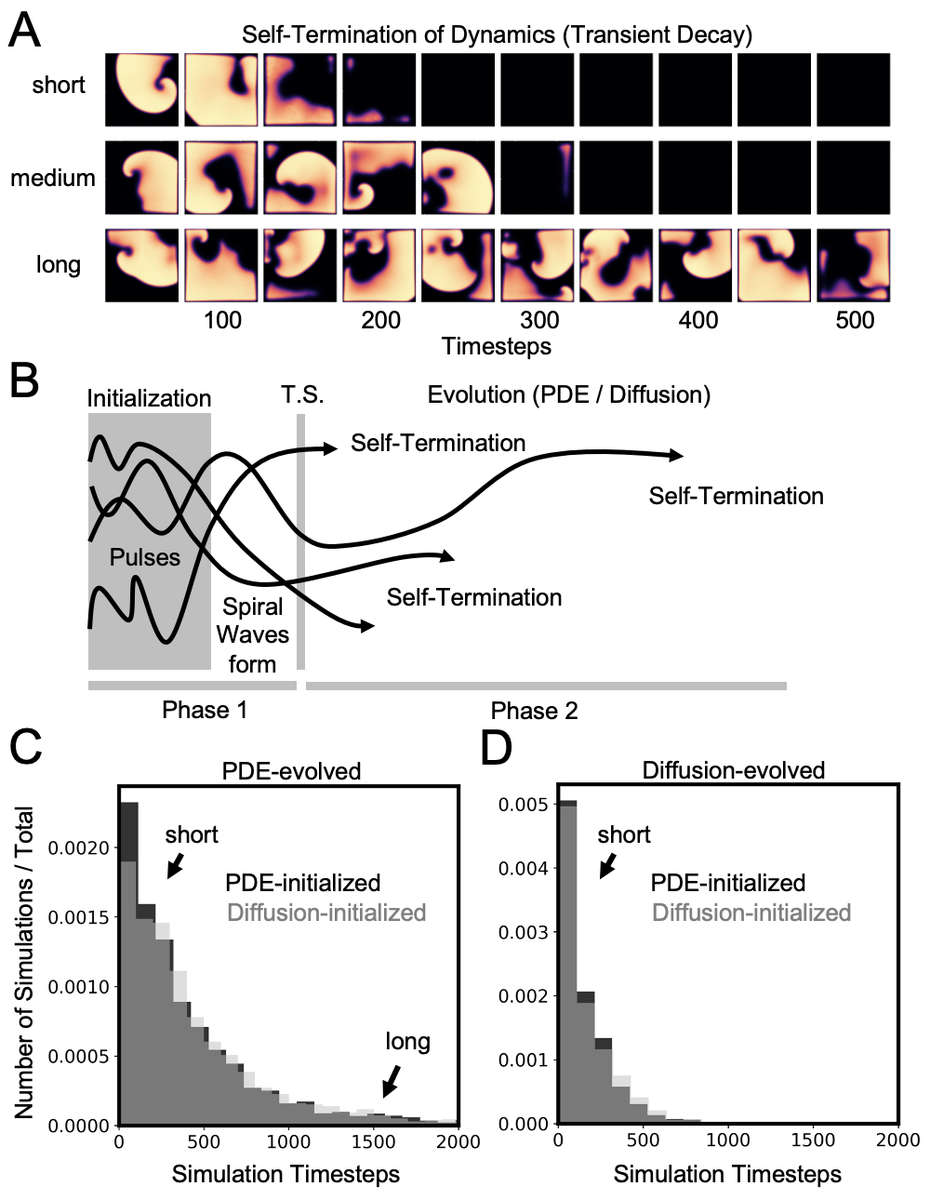}
  \caption{
{Self-termination characteristics of real and 'fake' spiral wave dynamics.
\textbf{A} Short-, medium- and long-lived spiral wave dynamics simulated with the biophysical model.
\textbf{B} Trajectories of different spiral wave episodes. In phase 1, the spiral waves are induced with a random pulse protocol using the biophysical model. Training samples (T.S.) were sampled over a period of $300$ frames right after the spiral waves had formed, see also Fig.~\ref{fig:2Dclassification}B). 
In phase 2, the spiral wave dynamics were further evolved using either the biophysical model or diffusion model from eq.~(\ref{eq:2Devolution}). 
At the beginning of phase 2, the simulations were either initialized or continued with the real dynamical states $(u,r)$ from phase 1 or the 'fake' dynamical states $(\tilde{u},\tilde{r})$ generated with the unconditional diffusion model from eq.~(\ref{eq:2Ddreaming}). 
Self-termination times were measured for all $4$ cases with respect to the beginning of phase 2.
\textbf{C,D} Distributions of self-termination (or survival) times measured over an ensemble of $5,000$ spiral wave dynamics showing an exponential decay. 
\textbf{C} The dynamics were evolved with the biophysical model (PDE) and initialized with either the biophysical or diffusion model.
\textbf{D} The dynamics were evolved with the diffusion model and initialized with either the biophysical or diffusion model.}
  }
  \label{fig:transients}
\end{figure}

The diffusion model used to generate the wave patterns in Fig.~\ref{fig:2Dclassification}B) was trained to generate two-dimensional spiral wave patterns including both dynamic variables from noise in an unconditional fashion (Task 6):
 \begin{eqnarray}
  \label{eq:2Ddreaming}
\xi(x,y) \rightarrow (\tilde{u},\tilde{r})(x, y)
\end{eqnarray}
The model was trained on spiral wave patterns obtained with the Aliev-Panfilov model with a fixed parameter set, see Table \ref{tab:parameters} (Task 6), which we adapted from \citet{Lilienkamp2017}.
Consequently, the model is not conditioned with input parameters, as in section \ref{sec:results:parameterspecific}, but can dream up any spiral wave pattern it can come up with given its training, see Fig.~\ref{fig:2Dclassification}B) and also Supplementary Videos 1 and 2.
Moreover, the model is completely unrestricted in that it is not trained to perform certain tasks, such as inpainting, nor constrained by certain boundary conditions or parameters that guide the generative process.
The unconditional model was trained with $50,000$ training samples of spiral wave patterns $(u,r)(x,y)$ simulated in an isotropic excitable medium with size $128 \times 128$ pixels, as shown in Fig.~\ref{fig:2Dclassification}A).
{All simulations ran for a fixed simulation time, until the end of phase 1 shown in Fig.~\ref{fig:transients}B).
They were initialized with a random pulse protocol\cite{Lilienkamp2017} to cause wave break and induce spiral wave dynamics, and they were stopped shortly after the spiral wave dynamics had fully developed, see also next section \ref{sec:results:transients}.
The training samples (T.S.) were sampled from the last $300$ time steps at the end of phase 1 of each simulation from an ensemble of $5,000$ simulations, see Fig.~\ref{fig:transients}B).
The images in Fig.~\ref{fig:2Dclassification} A,B) are a random selection from the simulated training samples and diffusion-generated data, respectively (16 images from each class).}

\subsection{{Self-Termination} Behavior of Diffusion-generated Spiral Waves}
\label{sec:results:transients}
Spiral wave dynamics eventually self-terminate if one waits long enough.
{Fig.~\ref{fig:transients}A) shows examples of short-, medium- and long-lived spiral wave dynamics which were simulated with the same parameters as in the previous section using the biophysical model from eqs.~(\ref{eq:modelu}--\ref{eq:modelr}).
Two of the three examples survive for only about $100$-$300$ simulation time steps before self-termination, while one survives for longer than $500$ simulation time steps.
When performing many simulations, most spiral wave dynamics self-terminate rather quickly, while only few episodes survive for long times, see also Fig.~\ref{fig:transients}B,C). 
The overall distribution of self-termination or survival times of spiral wave dynamics was previously found to be exponential\cite{Lilienkamp2017}.
Here, we found a similar behavior with diffusion-generated spiral wave dynamics.}

{We performed the same simulations as in the previous section \ref{sec:results:similarity}, but let the simulations run until they eventually self-terminated, see phase 2 in Fig.~\ref{fig:transients}B).
We also initiated simulations using the unconditional diffusion model (Task 6) from eq.~(\ref{eq:2Ddreaming}) from the previous section, since this model generates a full dynamical state $(u,r)(x, y)$, and also let those simulations run until they eventually self-terminated.
Both types of simulations were evolved using the biophysical model (PDE-evolved).
The histogram in Fig.~\ref{fig:transients}C) shows the distributions of self-termination times (or survival times) for ensembles of $5,000$ simulations of diffusion-initialized spiral wave dynamics (gray) vs. conventional spiral wave dynamics (black).
The self-termination times were calculated with respect to the beginning of phase 2.
Both distributions match, demonstrating that the diffusion-generated spiral wave patterns adhere to the same self-termination statistics with a similar exponential decay rate as their biophysical counterparts.
In other words, it appears that the unconditional diffusion-generated spiral wave patterns from Fig.~\ref{fig:2Dclassification}B) do not just look like the real spiral wave patterns shown in Fig.~\ref{fig:2Dclassification}A), but correspond to real physical solutions.}

{In another experiment, we compared the self-termination statistics when both spiral wave dynamics in phase 2 are evolved with the time-stepping diffusion model from eq.~(\ref{eq:2Devolution}) in section \ref{sec:results:evolve}.
The histogram in Fig.~\ref{fig:transients}D) shows that both distributions match (calculated for ensembles of $5,000$ simulations each).
However, both diffusion-evolved dynamics self-terminate much sooner than their PDE-evolved counterparts, highlighting that the time-stepping diffusion model from section \ref{sec:results:evolve} behaves differently than the finite differences time-stepping scheme of the biophysical model. 
This correlates with the observation in section \ref{sec:results:evolve} that diffusion-evolved spiral wave dynamics appear to exhibit less wave break than the corresponding PDE-evolved dynamics.
Less wave break could contribute to a shortening of the survival times of the dynamics.
Also, while measuring the self-termination times with the diffusion-evolved dynamics, we encountered one curious phenomenon: shortly before self-termination, the dynamics would be abruptly taken over by severe noise. 
This issue could be solved by adding more training data of self-termination events.
Taken together, these findings suggest that diffusion-generated spiral wave dynamics adhere to the same laws as their biophysical counterparts, but there are some reservations that require further exploration.}

\section{Discussion}
\label{sec:discussion}

Generative AI provides {the potential for} many promising applications in the biological and biomedical sciences \cite{}.
Here, we demonstrated that denoising diffusion probabilistic models (DDPMs) can be used to model waves of electrical excitation  in cardiac tissue.
Diffusion models can be used to reconstruct or create parameter-specific wave patterns, and, most importantly, simulate electrical wave propagation in a data-driven manner.
In other words, diffusion models can learn to evolve cardiac wave dynamics from previously seen data without knowledge about the underlying physics. 
Therefore, they could potentially be used to create a data-driven model of the heart's electrophysiological system from measurement data.
We found that diffusion models not only generate electrical wave dynamics that look like and are visually indistinguishable from simulated wave dynamics, but the diffusion-generated dynamics also preserve some of the inherent characteristics of the original dynamics. 
For instance, we found that diffusion-generated spiral wave dynamics exhibit very similar
self-termination statistics as their counterparts in excitable media, see section \ref{sec:results:transients}. 

While we have some confidence that the diffusion-generated waves are indeed legitimate solutions, we also remain cautious and further research is needed to confirm whether diffusion models provide a valid and robust alternative to conventional biophysical modeling. 
At this point, we cannot rule out that diffusion models merely emulate rather than simulate spiral wave dynamics.
This concern is particularly critical when the dynamics are chaotic and sensitive to slight physical perturbations or differences in the numerical integration.
We evolved both simpler and more complicated spiral wave dynamics, and while the diffusion model generated plausible-looking simpler spiral wave dynamics for very long times (in contrast to U-Net) which co-evolved over a reasonable period of time with the ground-truth dynamics (keep in mind that even different solvers would lead to diverging results), the more complicated spiral wave dynamics diverged very quickly from and exhibited less wavebreak than their ground-truth counterpart.
The latter observation could be an indication that bias in the training data influences the behavior of the dynamics in yet inexplicable ways.
{Additionally, we observed unfamiliar artifacts, such as a sudden onset of extreme noise shortly before the self-termination of spiral wave dynamics. This particular artifact could be mitigated by including more training data of self-termination events, which hints at fundamental issues with selective and insufficient training data.}

A major concern with diffusion models is their ability to hallucinate. 
Hallucination is an inherent property of generative modeling and a feature and bug at the same time. 
Diffusion models can generate a continuum of outputs of which some are made up and false.
The main issue is that the false output is hard to identify as diffusion models excel at learning the data distribution and generating realistically looking data points from this distribution.
This raises concerns over the applicability of diffusion models in healthcare, where they could produce misleading output which could lead to an incorrect diagnosis or treatment.
Here, we found that the extent to which diffusion models hallucinate is related to how much the task that the network is supposed to perform is constrained.
If the problem is more constrained, then the space of possible solutions becomes smaller and there is less potential for hallucination (e.g. when evolving dynamics).
Therefore, sufficiently constraining diffusion models as well as developing methods to quantify and mitigate hallucination is essential.
Nevertheless, the perceived weaknesses with regard to hallucination can also be a major advantage in other situations:
diffusion models excel when tasked to generate a starting point in underconstrained tasks, and therefore they could serve as a powerful prior for difficult cardiac modeling or reconstruction tasks.

Overall, despite the potential drawbacks, diffusion models are a promising tool with many potential applications in cardiac research and diagnostics.
Diffusion models can in principle generate parameter- or even model-specific (scroll) wave dynamics in three-dimensional heart-shaped geometries, suggesting that they could be used to simulate atrial or ventricular fibrillation in an individualized segmentation of a patient's heart while also integrating patient- or disease-specific {information (e.g. ion channel abnormalities)} and /or measurement data (e.g. catheter mapping or electrocardiographic data).
In particular, diffusion models offer the possibility to learn different integration time scales, perform simulations at arbitrary resolutions, and skip the tedious part of finding the right initial conditions for simulating such dynamics as they can readily generate spiral and scroll wave patterns instantaneously.

\section{Conclusions}
\label{sec:conclusions}

We demonstrated that denoising diffusion probabilistic models (DDPMs) can be used for generating electrophysiological wave patterns in cardiac tissue. 
They can be used for recovering missing data, evolving spatio-temporal dynamics, generating electrophysiological wave patterns in arbitrary geometries, or generating parameter-specific dynamics, among other tasks.
The diffusion-generated waves are visually indistinguishable from and behave very similar to waves simulated with biophysical models.
However, diffusion models tend to hallucinate with insufficient constraining, {produce artifacts in situations in which training data is lacking} and produce high computational upfront costs for training. 
In the future, diffusion models could be used for data-driven modeling of various physiological phenomena in the heart.

\bibliography{references}%

\section{Supplementary Material}
Additional figures and descriptions of the Supplementary Videos are provided in the Supplementary Material document. Supplementary Videos can be found at: \url{https://cardiacvision.ucsf.edu/videos/diffusion/}.

\section{Data Availability Statement}
All source codes and data generated in this study will be made available at \url{https://github.com/cardiacvision/diffusion} before publication.

\section{Funding}
This research was funded by the University of California, San Francisco, and the National Institutes of Health (DP2HL168071). The RTX A5000 GPUs used in this study were donated by the NVIDIA Corporation via the Academic Hardware Grant Program (to JL and JC).

\section{Author Contributions}
TB  developed the deep learning methodology and performed the data analysis.
TB and JC performed the two-dimensional simulations.
JC aided with the data analysis.
JL performed the SPH and bulk simulations.
TB and JC designed the figures.
JC wrote the manuscript.
TB aided in writing the manuscript.
All authors discussed and interpreted the results and read and approved the final version of the manuscript.
JL and JC conceived the research.

\section{Conflict of Interest}
The authors declare that the research was conducted in the absence of any commercial or financial relationships that could be construed as a potential conflict of interest.

\end{document}